\documentclass[twocolumn,superscriptaddress,showpacs,preprintnumbers,amsmath,amssymb, pre]{revtex4}
\usepackage{graphicx}
\usepackage{bm}
\usepackage{lipsum}

\begin{document}

\title{Active acoustic switches using 2D granular crystals}

\author{Qikai Wu}
\affiliation{Department of Mechanical Engineering and Materials Science, Yale University, New Haven, Connecticut, 06520, USA}
\author{Chunyang Cui} 
\affiliation{Department of Water Resource Science and Engineering, Tsinghua University, Beijing, China}
\author{Thibault Bertrand}
\affiliation{Department of Mathematics, Imperial College London, South 
Kensington Campus, London SW7 2AZ, England, UK}
\affiliation{Department of Mechanical Engineering and Materials Science, Yale University, New Haven, Connecticut, 06520, USA}
\author{Mark D. Shattuck}
\affiliation{Department of Physics and Benjamin Levich Institute, The City College of the City University of New York, New York, 10031, USA}
\affiliation{Department of Mechanical Engineering and Materials Science, Yale University, New Haven, Connecticut, 06520, USA}
\author{Corey S. O'Hern}
\affiliation{Department of Mechanical Engineering and Materials Science, Yale University, New Haven, Connecticut, 06520, USA}
\affiliation{Department of Physics, Yale University, New Haven, Connecticut, 06520, USA}
\affiliation{Department of Applied Physics, Yale University, New Haven, Connecticut, 06520, USA}

\date{\today}

\begin{abstract}
We employ numerical simulations to study active transistor-like
switches made from two-dimensional (2D) granular crystals containing
two types of grains with the same size, but different masses. We tune
the mass contrast and arrangement of the grains to maximize the width
of the frequency band gap in the device.  The input signal is applied
to a single grain on one side of the device, and the output signal is
measured from another grain on the other side of the device.  Changing
the size of one or many grains tunes the pressure, which controls the
vibrational response of the device. Switching between the on and off
states is achieved using two mechanisms: 1) pressure-induced switching
where the interparticle contact network is the same in the on and off
states, and 2) switching through contact breaking. In general, the
performance of the acoustic switch, as captured by the gain ratio and
switching time between the on and off states, is better for
pressure-induced switching.  We show that in these acoustic switches
the gain ratio between the on and off states can be larger than $10^4$
and the switching time (multiplied by the driving frequency) is
comparable to that obtained recently for sonic crystals and less than
that for photonic transistor-like switches. Since the self-assembly of
grains with different masses into 2D granular crystals is challenging,
we describe simulations of circular grains with small circular knobs
placed symmetrically around the perimeter mixed with circular grains
without knobs.  Using umbrella sampling techniques, we show that
devices with grains with $3$ knobs most efficiently form
the hexagonal crystals that yield the largest
band gap.
\end{abstract}

\maketitle

\section{Introduction}
\label{intro}

A number of recent studies have demonstrated the potential for
granular crystals to serve as switches~\cite{alagoz},
rectifiers~\cite{boechler}, and other logic elements~\cite{li} in
circuits that use mechanical rather than electrical signals. These
mechanical devices have potential applications in vibration
isolation~\cite{gantzounis}, acoustic cloaks~\cite{zigoneanu}, and
one-way sound propagation~\cite{cummer}. Many prior studies have used
one-dimensional (1D) granular chains as model systems~\cite{boechler2}
and relied on the nonlinear Hertzian interparticle contact law to
tailor the acoustic response~\cite{sokolow,nesterenko,schreck1}. For
example in Ref.~\cite{li}, researchers developed an acoustic switch by
taking advantage of the fact that 1D granular chains composed of steel
beads possess a high-frequency cutoff $\omega_{\rm max}$, beyond which
an input signal cannot propagate. Thus, when the system is driven at
$\omega_0 > \omega_{\rm max}$, the response is extremely small,
i.e. it exists in the ``off" state. However, when the system is also
driven at frequency $\omega_c < \omega_{\rm max}$, nonlinearities from
the Hertzian interactions between grains can induce a strong response
at $\omega_0$ (i.e. produce an ``on" state), as well as linear
combinations of $\omega_0$ and $\omega_c$.  The authors showed that
the amplitude of the response at $\omega_0$ in the on state was $3.5$
orders of magnitude larger than that of the off state~\cite{li}. This
seminal work demonstrated the ability to actively control mechanical
signal propagation in 1D granular chains.

\begin{figure}[h!]
\includegraphics[width=3in]{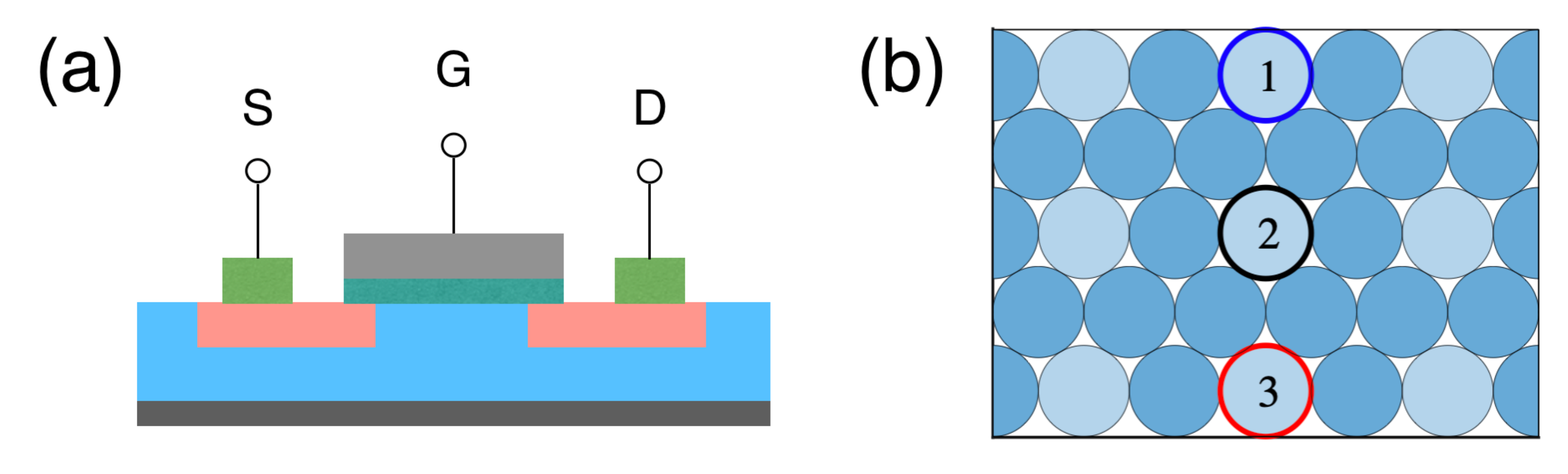}
\caption{(a) A schematic of a metal-oxide-semiconductor field-effect 
transistor (MOSFET) with gate (G), source (S), and drain (D) ports and 
(b) a schematic of a switch made from a 2D granular crystal with three 
ports for the (1) output, (2) control, and (3) input signals.}
\label{transistor}
\end{figure}

Transistors are fundamental components of modern electrical devices
that perform logic operations by amplifying or switching electrical
signals~\cite{bardeen}. In this study, we numerically design a
transistor-like acoustic switch using 2D granular crystals composed of
grains with two different masses $m_L$ and $m_S$. In a typical
field-effect transistor, the drain-to-source current is controlled by
the voltage applied between the gate and source
terminals. Analogously, in our system, the mechanical response will be
controlled by the applied pressure. As shown in Fig.~\ref{transistor},
we will consider three-port devices. We will send mechanical signals
to a single particle (port $3$) on one side of the system, apply
pressure by changing the size of a single or many grains (port $2$),
and measure the power spectrum of the displacements of another grain
on the other side of the system (port $1$).

\begin{figure*}
\includegraphics[width=1\textwidth]{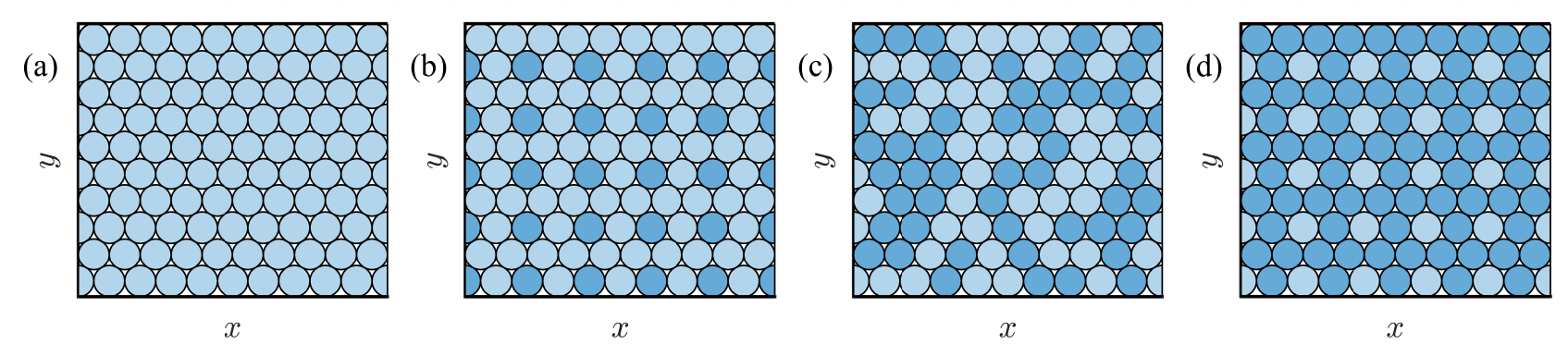}
\caption{Mechanically stable packings of $N=100$ disks with the same
size, two different masses, and mass ratio $m_L/m_S=10$ arranged on
a hexagonal lattice with periodic and fixed boundary conditions in 
the $x$- and $y$-directions, respectively. In (a), the system is homogeneous with $N_L= 0$
(dark blue) and $N_S=N$ (light blue). In (b), we set $N_L=25$ and
$N_S=75$.  The first row contains all small masses. In the second
row, the large and small masses alternate. The third row alternates
between large and small masses, and this order repeats for a total
of ten rows. In (c), $N_L=50$ and $N_S=50$ and large and small
masses are distributed randomly on the hexagonal lattice. Panel (d) is similar
to (b), except inverted with $N_L=75$ and $N_S=25$.}
\label{EigenFreq}
\end{figure*}

Granular crystals composed of two types of grains with the same size,
but with mass contrast $m_L/m_S >1$, possess band gaps in their
vibrational density of states~\cite{boechler2,goncu2}. The width of
the band gap depends strongly on pressure~\cite{goncu1}. Thus, by
varying the pressure at fixed driving frequency, we can change the
range of the frequency band gap so that the driving frequency occurs
within or outside the band gap. When the system is excited at a
frequency within the band gap, the signal will not propagate and the
switch is off. When the system is excited at a frequency outside the
band gap, it will propagate and the switch is on. Thus, by changing
the pressure, we can actively switch the device between the off and on
states.  In addition, using 2D granular crystals allows us to
determine the effects of the polarization of the mechanical signal and
contact breaking~\cite{tournat,schreck2}, where grains come in and out
of contact during vibration, on the performance of acoustic switches.

We will quantify the performance of the acoustic switch by measuring
its gain, which is the ratio of the amplitude of the displacement
spectrum at the driving frequency for the output versus that of the
input particle (via ports $1$ and $3$). We find that the ratio of the
gain for the on and off states of the device can be four orders of
magnitude or larger. We also characterize the time required to switch
between the on and off states and vice versa. We find that there is a
trade-off between the switching time and gain ratio. We achieve the
fastest switching times for devices with the smallest gain ratios
between the on and off states. In addition, we investigated the effect
of contact breaking on the performance of granular acoustic switches.
We find that when changes in pressure cause contact breaking in the
device, the performance of the switch is degraded.  In particular,
devices with contact breaking can only achieve modest gain ratios,
where the gain for the on state is $1.5$ orders of magnitude larger
than that for the off state.  We also studied the performance of the
acoustic switch when we adjust the sizes of a single versus multiple
grains to induce changes in pressure. Adjusting the sizes of multiple
grains allows the device to achieve larger gain ratios. In addition,
since it is typically difficult to generate 2D granular crystals
containing grains with different masses in both simulations and
experiments~\cite{saadatfar,reitz}, we also describe a method to
generate granular crystals in 2D using circular grains that include
small circular knobs on their surfaces. Using molecular dynamics
simulations with advanced sampling techniques allows us to determine 
the number and placement of the knobs that yield the most efficient 2D
crystallization.

This article includes three additional sections and two Appendices.
In the Methods section, we describe calculations of the vibrational
density of states for 2D granular crystals composed of two types of
grains with the same size, but with mass contrast $m_L/m_S >1$. We
measure the width of the frequency band gap as a function of the mass
contrast, arrangement of the heavy and light grains, and pressure.
In addition, we describe the input
signal, how the output signal will be measured, and the methods that
will be used to change the pressure in the device. In the Results
section, we show our calculations of the gain ratios for the on and
off states in devices where the pressure is varied and in regimes
where the network of interparticle contacts is fixed or fluctuates. We
provide results for the gain ratios for systems in steady state, and
study the gain as a function of time after the device switches from on
to off and vice versa. We also describe molecular dynamics simulations
coupled with advanced sampling methods in 2D of circular grains
containing small circular knobs on their surfaces and identify the
number and placement of knobs that give rise to the most efficient
crystallization.  In the final section, we summarize our most
important results, suggest future calculations, and discuss the
possibility to build mechanical circuits that can perform logical
operations. The two Appendices provide additional technical details
that support the methods and results in the main text. In Appendix A,
we show that the numerical methods used to calculate the discrete
Fourier transform of the input and output signals do not affect our
results. In Appendix B, we show results for the performance of 2D
granular acoustic switches with small band gaps.

\section{Methods}
\label{methods}

To narrow the parameter space, we focus on 2D granular systems
composed of frictionless circular disks in the absence of gravity.
For most studies, the systems include two types of disks with the same
diameter $\sigma$, but different masses, $m_L$ and $m_S$, with $m_L >
m_S$. The $N=N_L+N_S$ disks (where $N_L$ and $N_S$ are the numbers of
disks with mass $m_L$ and $m_S$, respectively) interact via the
pairwise, purely repulsive linear spring potential,
\begin{equation}
\label{Ep}
U(r_{ij})=\frac{\epsilon}{2} \left(1-\frac{r_{ij}}{\sigma}\right)^{2}\Theta\left(1-\frac{r_{ij}}{\sigma}\right),
\end{equation}
where $r_{ij}$ is the separation between the centers of disks $i$ and
$j$, $\epsilon$ is the energy scale of the repulsive interaction, and
$\Theta(x)$ is the Heaviside step function that sets $U(r_{ij})=0$
when the disks are not in contact with $r_{ij} > \sigma_{ij}$.  For
most studies, the simulation cell is rectangular with area $A=L_x L_y$
and dimensions $L_x= N_x \sigma$, and $L_y= N_y \sqrt{3}\sigma/2$,
where $N=N_x N_y$, and $N_x$ and $N_y$ are the number of particles in
the $x$- and $y$-directions, so that it can accommodate a hexagonal
lattice. We implement periodic boundary conditions in the
$x$-direction, and fixed, flat boundaries in the
$y$-direction. Interactions between a circular grain and the wall
are implemented by assuming that a ghost particle is placed at a
symmetric position behind the wall. We focus on systems with
relatively small $N$, from $N=30$ to $100$ grains, since it is
difficult to self-assemble perfect crystalline structures in large
systems~\cite{saadatfar,reitz}. Below, lengths, energies, stresses,
and frequencies will be given in units of $\sigma$, $\epsilon$,
$\epsilon/\sigma^2$, and $\sqrt{\epsilon/m_S\sigma^2}$, respectively.

Most of the systems we consider are mechanically stable with a full
spectrum of $2N$ nonzero vibrational frequencies, $\omega_k$, with
$k=1$,$\ldots$,$2N$. The vibrational frequencies are obtained by
calculating the eigenvalues $\lambda_k = \omega^2_k$ of the
mass-weighted dynamical matrix~\cite{tanguy} ${\cal M}_{kj}=M^{-1}_{ki}H_{ij}$, where
$H_{ij} = \partial^2 U/\partial \xi_i \partial \xi_j$ is the Hessian
of the total potential energy $U=\sum_{i>j} U(r_{ij})$, $\xi_i=x_i$,
$y_i$, and $M_{ij}=m_{L,S}\delta_{ij}$ is the diagonal mass matrix.
We also determine the eigenvectors ${\vec \lambda}^k$ that correspond to 
each eigenfrequency $\omega_k$ with ${\vec \lambda}^k \cdot {\vec \lambda}^k =1$, 
where ${\vec \lambda}^k=\{x^k_1, y^k_1,\ldots,x^k_N, y^k_N\}$.  

\begin{figure}[h!]
\includegraphics[width=3in]{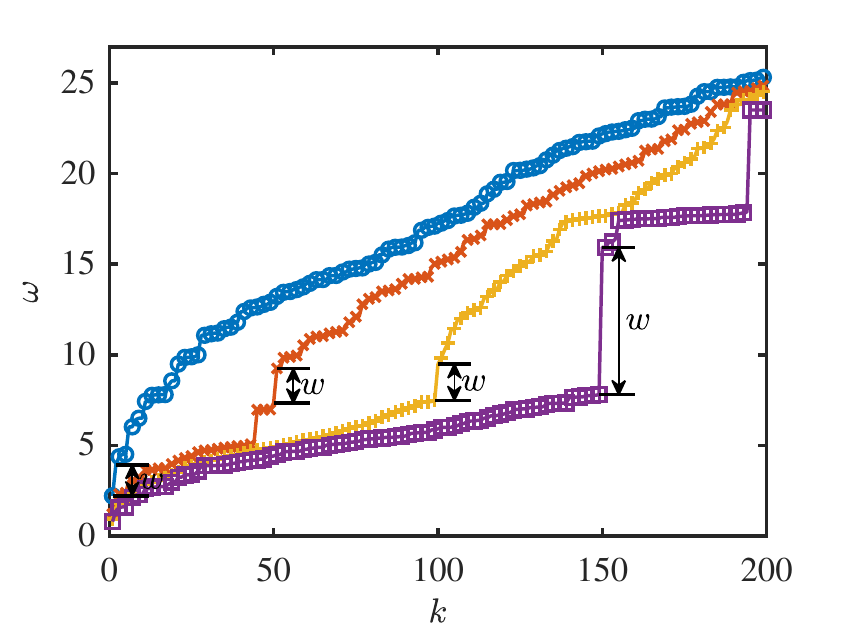}
\caption{Eigenfrequencies of 
the mass-weighted dynamical matrix $\omega_k$, sorted in ascending order
and indexed by $k$, for the $N=100$
configurations in Fig.~\ref{EigenFreq} (a) circles, (b) exes, (c) plusses, 
and (d) squares with periodic and fixed boundary conditions in the $x$- and 
$y$-directions, respectively. $w$ 
indicates the maximum band gap in the eigenfrequency spectrum.}
\label{EigenFreq_2}
\end{figure}

We calculate the eigenfrequency spectrum of the mass-weighted
dynamical matrix for several arrangements of the large and small
masses on a hexagonal lattice with $N=100$ shown in
Fig.~\ref{EigenFreq}. We illustrate in Fig.~\ref{EigenFreq_2} that for
a hexagonal lattice with a uniform mass distribution
[Fig.~\ref{EigenFreq}~(a)], the frequency spectrum is nearly
continuous with a high frequency cutoff $\omega_{\rm max} \approx
25$. For mixtures of large and small masses with a mass ratio $m_L/m_S
=10$ [Fig.~\ref{EigenFreq} (b) and (c)], a small frequency band gap
develops in the range $5 \lesssim \omega \lesssim 8$. For each
eigenfrequency spectrum, we identify the maximum frequency difference $w=\max_k
(\omega_{k+1}-\omega_k)$.

\begin{figure}[h!]
\includegraphics[width=3in]{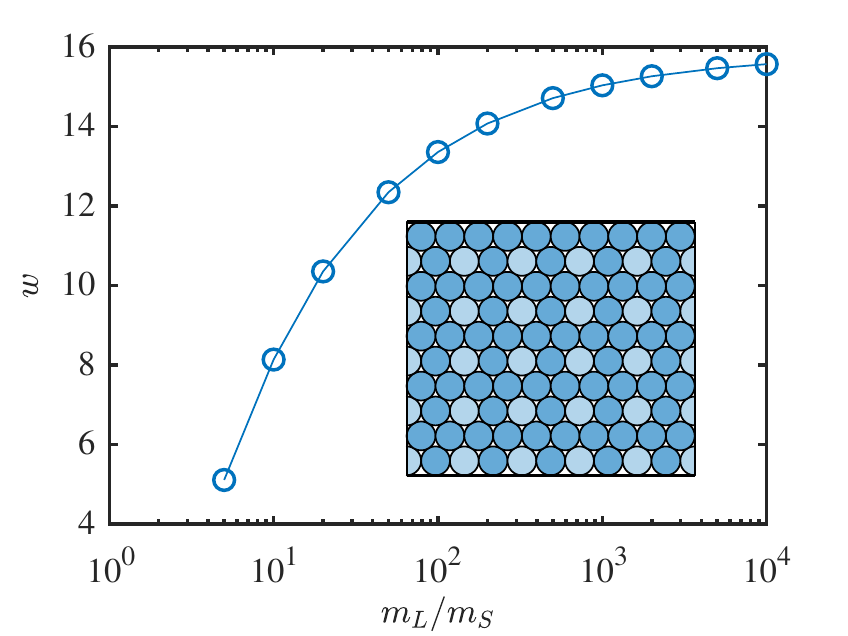}
\caption{The width $w$ of the maximum band gap in the eigenfrequency spectrum of the mass-weighted dynamical 
matrix for the configuration in Fig.~\ref{EigenFreq} (d) (and inset)
as a function of the mass ratio $m_L/m_S$.}
\label{BandGap}
\end{figure}

We find that the arrangement of large and small masses that gives rise
to the largest band gap $w$ is the alternating pattern in
Fig.~\ref{EigenFreq} (d). In Fig.~\ref{BandGap}, we show that for the
optimal arrangement of large and small masses
(i.e. Fig.~\ref{EigenFreq} (d)), $w$ increases with $m_L/m_S$,
reaching a plateau of $w \approx 16$ in the $m_L/m_S \rightarrow
\infty$ limit.  For most of our studies, we use a mass ratio,
$m_L/m_S=10$, with $w\approx 10$.

\begin{figure}[h!]
\includegraphics[width=3in]{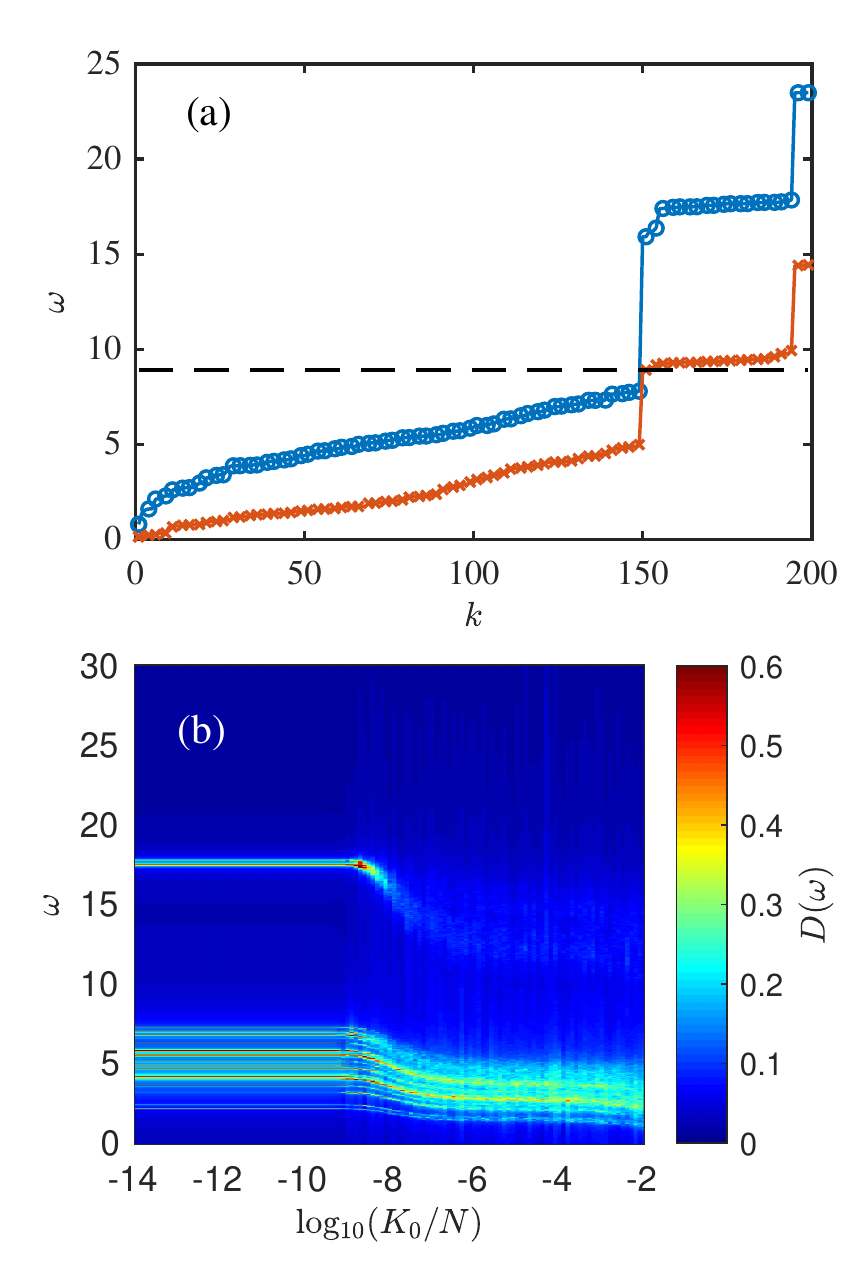}
\caption{(a) Spectrum of eigenfrequencies of the mass-weighted 
dynamical matrix sorted in ascending order with index $k$ for  
systems with $N=100$ disks, $m_L/m_S=10$, and arranged on a hexagonal 
lattice in the optimal configuration in Fig.~\ref{EigenFreq} (d) at 
pressure $p=10^{-3}$ (circles) and $1$ (exes). The dashed line 
indicates a driving frequency at which the acoustic switch can operate. 
(b) The Fourier transform of the velocity correlation function $D(\omega)$
for the mechanically stable packing in Fig.~\ref{EigenFreq} (d) at 
pressure $p=10^{-4}$ after adding velocities to all grains such that 
the eigenfrequencies of the mass-weighted 
dynamical matrix are included with equi-partition
of the total kinetic energy $K_0$. The color scale from dark red to violet represents decreasing
$D(\omega)$ on a linear scale.}
\label{vCorr}
\end{figure}

The width of the frequency band gap can also be tuned by changing
the pressure of the system. When all of the disks are at contact and
placed on a hexagonal lattice, the packing fraction is $\phi_{\rm
  xtal} = \pi/2\sqrt{3} \approx 0.91$ for systems with periodic
boundary conditions in both the $x$- and $y$-directions (and $\approx
0.89$ for systems with fixed boundaries in the $y$-direction and
periodic boundaries in the $x$-direction), and the pressure $p= A^{-1}
\sum_{i>j} {\vec f}_{ij} \cdot {\vec r}_{ij}/2$ is nearly zero, where
${\vec f}_{ij} = -dU/d{\vec r}_{ij}$ is the repulsive force on disk
$i$ arising from disk $j$.  We can change the pressure of the system
by increasing or decreasing the diameter of the disks by an increment
in packing fraction $\Delta \sigma/\sigma = \Delta \phi/\phi$, or
equivalently by bringing the fixed walls in the $y$-direction closer
together or further apart.  We define the packing fraction as $\phi =
A^{-1} \sum_{i=1}^N \pi \sigma_i^2/4$, even for systems in which the
grains overlap. In Fig.~\ref{vCorr} (a), we show the spectrum of
eigenfrequencies of the mass-weighted dynamical matrix for the
configuration in Fig.~\ref{EigenFreq} (d) with $m_L/m_S = 10$ at low
$p=10^{-3}$ and high pressure $p=1$. For the system at low pressure,
we can set the driving frequency at $\omega_0 \approx 9$ in the band
gap, and the system exists in the off state. When we compress the
system to high pressure, all of the eigenfrequencies decrease, and the width 
of the band gap also decreases.  At high pressure, the driving frequency is no
longer in the band gap, and the system exists in the on state. Thus,
2D granular crystals can be switched from on to off and vice versa by
changing the pressure.

Contact breaking, a significant source of nonlinearity in granular
materials~\cite{schreck2,bertrand,qikai}, can also be used to switch
between the on and off states and vice versa in 2D granular crystals.
Contact breaking occurs when the system is driven at sufficiently
large amplitudes (e.g. through vibration or shear) so that the network
of interparticle contacts changes. The characteristic driving
amplitude at which contact breaking occurs decreases with pressure.
When the system can break interparticle contacts and form new ones,
the frequencies of strong peaks in the Fourier transform of the
velocity autocorrelation function of the disks will differ from the spectrum of
eigenfrequencies of the mass-weighted dynamical matrix.

To illustrate contact breaking and its effect on the vibrational response, we excite a 2D granular crystal by
setting the velocities of the grains such that all eigenmodes of the
mass-weighted dynamical matrix are included with equi-partition of the
total kinetic energy, $K_0$. To determine the vibrational response, we
calculate the Fourier transform of the normalized velocity
autocorrelation function,
\begin{equation}
\label{vCorr_equ}
D(\omega) = \int_0^\infty dt \frac{\langle \vec{v}(t_0+t) \cdot \vec{v}(t_0)\rangle}{\langle \vec{v}(t_0) \cdot \vec{v}(t_0)\rangle}e^{i\omega t},
\end{equation} 
where $\langle.\rangle$ indicates an average over all of the disks and
time origins $t_0$. In Fig.~\ref{vCorr} (b), we show $D(\omega)$ as a
function of $K_0/N$ for the optimal configuration in
Fig.~\ref{EigenFreq} (d) at $p=10^{-4}$. At small vibration amplitudes,
$D(\omega)$ is large at all of the $2N$ eigenfrequencies of the
mass-weighted dynamical matrix. When the vibration amplitude exceeds
$K_0/N \approx 10^{-9}$ existing contacts begin to break and new
contacts begin to form, $D(\omega)$ broadens and spreads to lower
frequencies. In particular, for amplitudes above $10^{-9}$, there is a
very weak response at high frequencies. Thus, contact breaking can
also be used to switch between the on and off states. For example,
when the system is driven at $\omega_0 =18$ at small $K_0/N$, the
switch is on. However, when the system is driven at the same $\omega_0$ with
amplitude $K_0/N \gtrsim 10^{-9}$, the switch is off.

\begin{figure}[h!]
\includegraphics[width=3in]{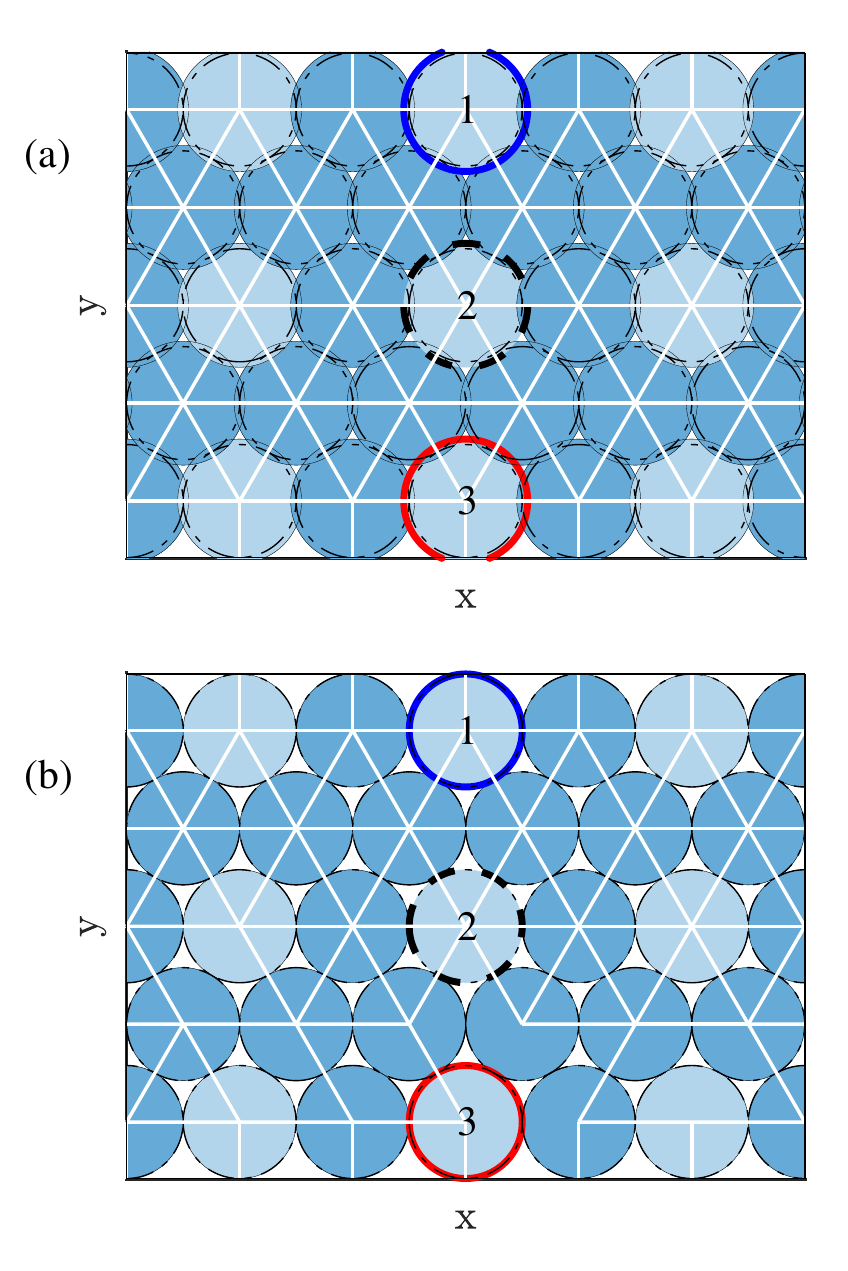}
\caption{(a) An illustration of a three port acoustic switch with fixed, flat
boundary conditions in the $y$-direction and periodic boundary
conditions in the $x$-direction.  The device includes $N=30$ disks
(with $N_L=21$ (dark), $N_S=9$ (light), and $m_L/m_S=10$) arranged 
on a hexagonal lattice. The solid white lines indicate 
the $N_c=90$ distinct contacts between disks. Disk $3$ is the input port, 
indicating 
where the system will be driven. The gain of the system is measured 
via the output port, labelled disk $1$.  The switch can be turned on and 
off by varying the pressure of the system through port $2$, e.g. by 
changing the size of a single disk or 
all disks in the system. Here, the device changes from pressure $p=10^{-6}$
(dashed outline) to $10^{-1}$ (solid outline) when all disks increase in 
size. (b) Illustration of the device 
in (a) at $p=10^{-6}$ with disk $3$ driven at $A_0=10^{-6}$ and frequency 
$\omega_0=16.0$, which causes contact breaking. In this snapshot, the 
device has four fewer contacts than in (a). The central grain with the 
dashed outline provides the pressure control when we use single-particle 
control for port $2$.}
\label{ThreePort}
\end{figure}

For the specific device geometry, we consider a three-port switch
built from the 2D granular crystal shown in
Fig.~\ref{ThreePort}. We will add sinusoidal
displacements with amplitude $A_0$ at driving frequency $\omega_0$ to a single disk on the
bottom wall (port $3$),
\begin{equation}
\label{drive}
x_3(t) = x_3^0 + A_0 \sin(\omega_0 t),
\end{equation}
where $x_3^0$ is the position of disk $3$ in the mechanically stable
packing.  When we add a continuous input signal, we also include a
viscous damping force for each disk $i$, ${\vec F}_i = -b {\vec v}_i$,
where $b$ is the damping coefficient. After the system reaches a
steady state, we determine the response of the system by measuring the
Fourier transform of the $x$-displacement of disk $1$ that is several
layers away from disk $3$ in the top wall (port $1$): $F_1(\omega) =
\int_0^{\infty} [x_1(t)-x_1^0] e^{i\omega t} dt$. The Fourier
transform is calculated numerically as discussed in
Appendix~A. The gain of the system is defined as the ratio of the
response at the output port $1$ to strength of the signal at the input
port $3$ at the driving frequency $\omega_0$:
\begin{equation}
\label{Gain}
G(\omega_0) = \frac{F_1(\omega_0)}{F_3(\omega_0)}.
\end{equation}  
Note that we chose the input and output signals to be in the
$x$-direction, which we assume has a significant overlap with the
eigenmodes of the system. We deliberately did not consider input and 
output signals along eigenmodes since they are difficult to measure 
experimentally in 2D granular media. 

We will actively control the response of the device (i.e. through port
$2$) by varying the pressure in the device. We will adjust the
pressure by changing the size of grain $i$:
$\Delta_i(t)=(\sigma_i(t)-\sigma)/\sigma$, where $\sigma$ is the
unperturbed diameter of the grains.  For the control signal, we can
also vary the fraction of grains $f$ whose sizes are changed by
$\Delta$.  Below, we will consider the extremes $f=1/N$ (one grain) and
$1$ (all grains). The case $f=1$ is depicted in Fig.~\ref{ThreePort}
(a). 

\section{Results}
\label{results}

We describe the results on acoustic switches constructed from 2D
granular crystals in four subsections.  In
Sec.~\ref{pressure_switching}, we focus on acoustic devices that can
switch between the on and off states by changing the size of all
particles in the system to control the pressure, and both the on and
off states have the same network of interparticle contacts.  These
devices can achieve large gain ratios of at least four orders of
magnitude between the on and off states. However, the switching times
are rather large, exceeding hundreds of oscillations of the driving
frequency.  Further, there is a tradeoff between gain ratios and
switching times, i.e.  the largest gain ratios are achieved for the
slowest switching times.  In Sec.~\ref{contact_breaking}, we discuss
acoustic devices in which contact breaking occurs, i.e., the on and
off states possess different interparticle contact networks. In
general, these devices have worse performance (smaller gain ratios)
than those for which the interparticle contact networks are the same
in the on and off states. However, switching between the on and off
states in these devices can be achieved at much lower pressures. In
Sec.~\ref{single_particle}, we discuss the pressure operating regime
for the acoustic device when the size of only a single control
particle is used to tune between the on and off states. In general, devices 
with a single control particle possess smaller gain ratios than those 
with many control particles. In
Sec.~\ref{periodic}, we describe a novel simulation technique, where
we add small circular knobs to the surface of circular grains, that can
robustly generate the ordered disk packing with the optimal
arrangement of more and less massive grains in Fig.~\ref{EigenFreq}
(d).  Similar techniques can be used in experiments to generate 2D 
granular crystals.

\subsection{Pressure-induced switching}
\label{pressure_switching}

\begin{figure}[h!]
\includegraphics[width=3in]{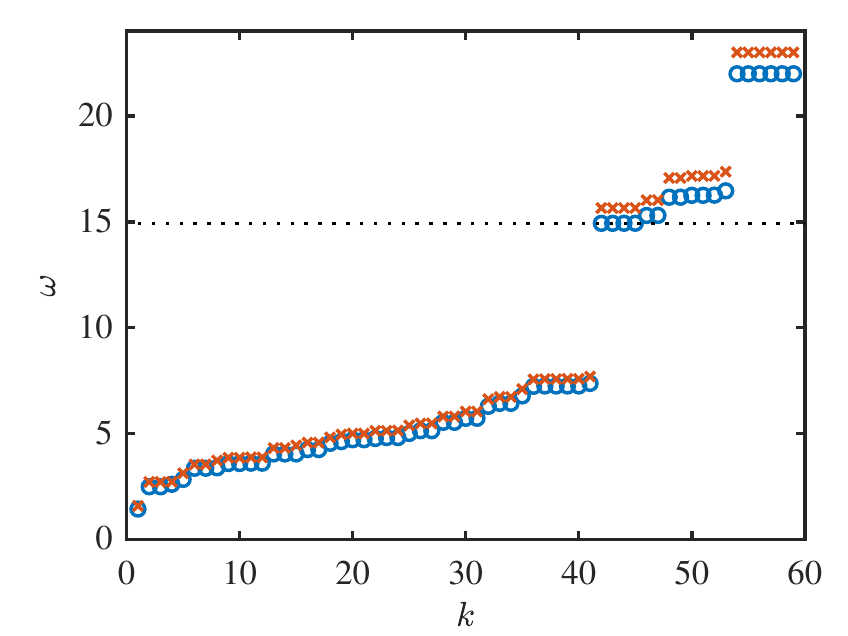}
\caption{The eigenfrequencies of the mass-weighted dynamical matrix
plotted in increasing order with index $k$ for the acoustic device in
Fig.~\ref{ThreePort} at pressure $p = 10^{-1}$ (the ``on'' state,
circles) and $3.2 \times 10^{-2}$ (the ``off'' state, exes). The
horizontal line at $\omega = 14.9$ indicates a potential driving
frequency that yields a large gain ratio between the on and off 
states.}
\label{AllPart_w}
\end{figure}

In Fig.~\ref{AllPart_w}, we show the eigenfrequencies of the
mass-weighted dynamical matrix for the device in Fig.~\ref{ThreePort} (a)
in the high pressure regime, $p=10^{-1}$ and $3.2\times
10^{-2}$. Changes in the pressure of the device allow us to tune the
frequency range of the band gap. When we drive the system at $\omega_0
=14.9$ with $p=10^{-1}$, we expect the gain to be large since the density of states
has weight at the driving frequency. In contrast, when we drive the
system at the same frequency and $p=3.2\times 10^{-2}$, there is no
weight in the density of states at the driving frequency and we expect the gain
be much smaller, even though the interparticle contact network is the same as that 
for the device at $p=10^{-1}$. 

\begin{figure}[h!]
\includegraphics[width=3in]{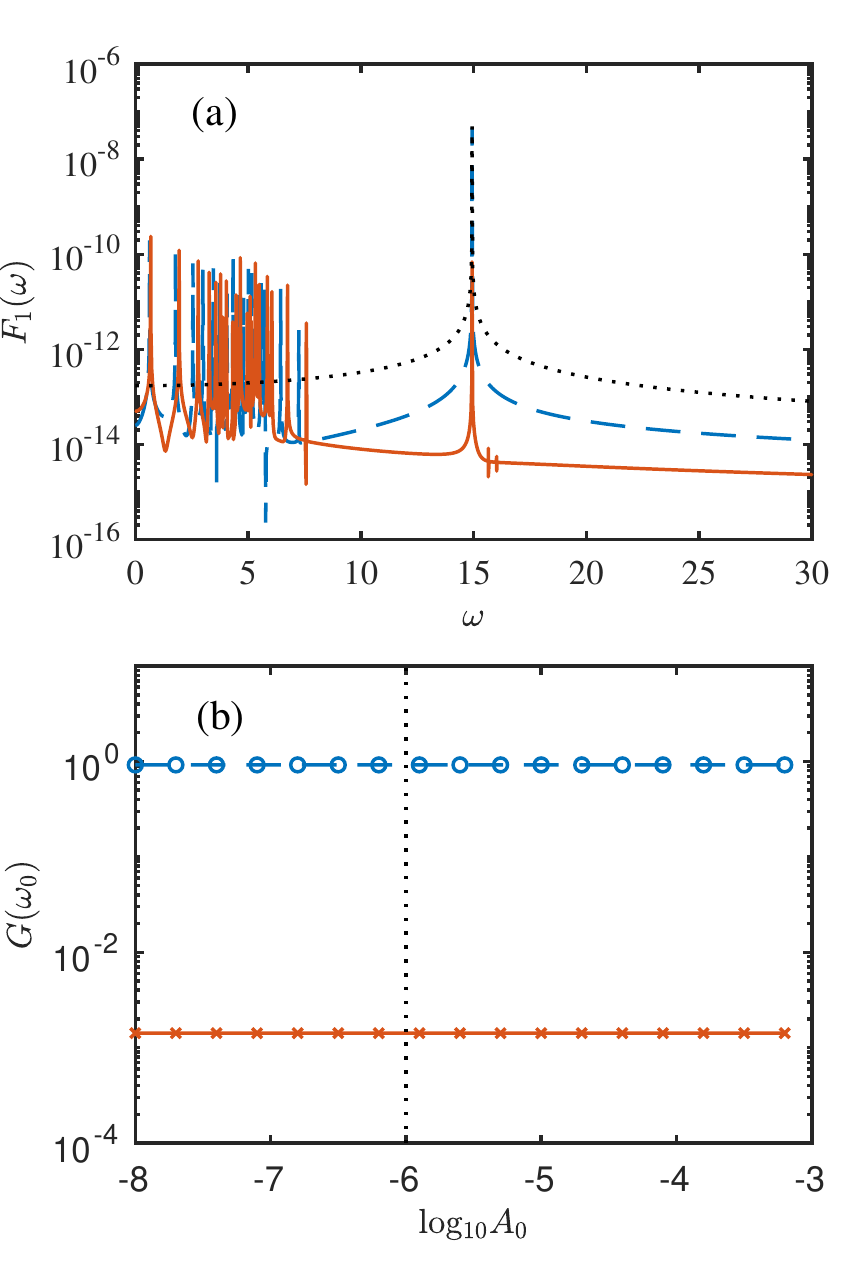}
\caption{(a) The Fourier transform $F_1(\omega)$ of 
the $x$-displacement of disk $1$ for the acoustic device with pressure $p = 10^{-1}$ 
(solid line) and $3.2 \times 10^{-2}$ (dashed line) obtained 
by driving disk $3$ sinusoidally with amplitude $A_0=10^{-6}$ 
and frequency $\omega_0=14.9$. The dotted line shows the Fourier transform 
$F_3(\omega)$ of the $x$-displacement of the input disk $3$. (b) The gain G($\omega_0$) (defined in Eq.~\ref{Gain}) plotted 
as a function 
of the driving amplitude $A_0$ with driving frequency $\omega_0=14.9$ 
for the device at pressure $p = 10^{-1}$ (open circles) and $3.2 \times 
10^{-2}$ (exes).}
\label{AllPart_Gain}
\end{figure}

\begin{figure}[h!]
\includegraphics[width=2.95in]{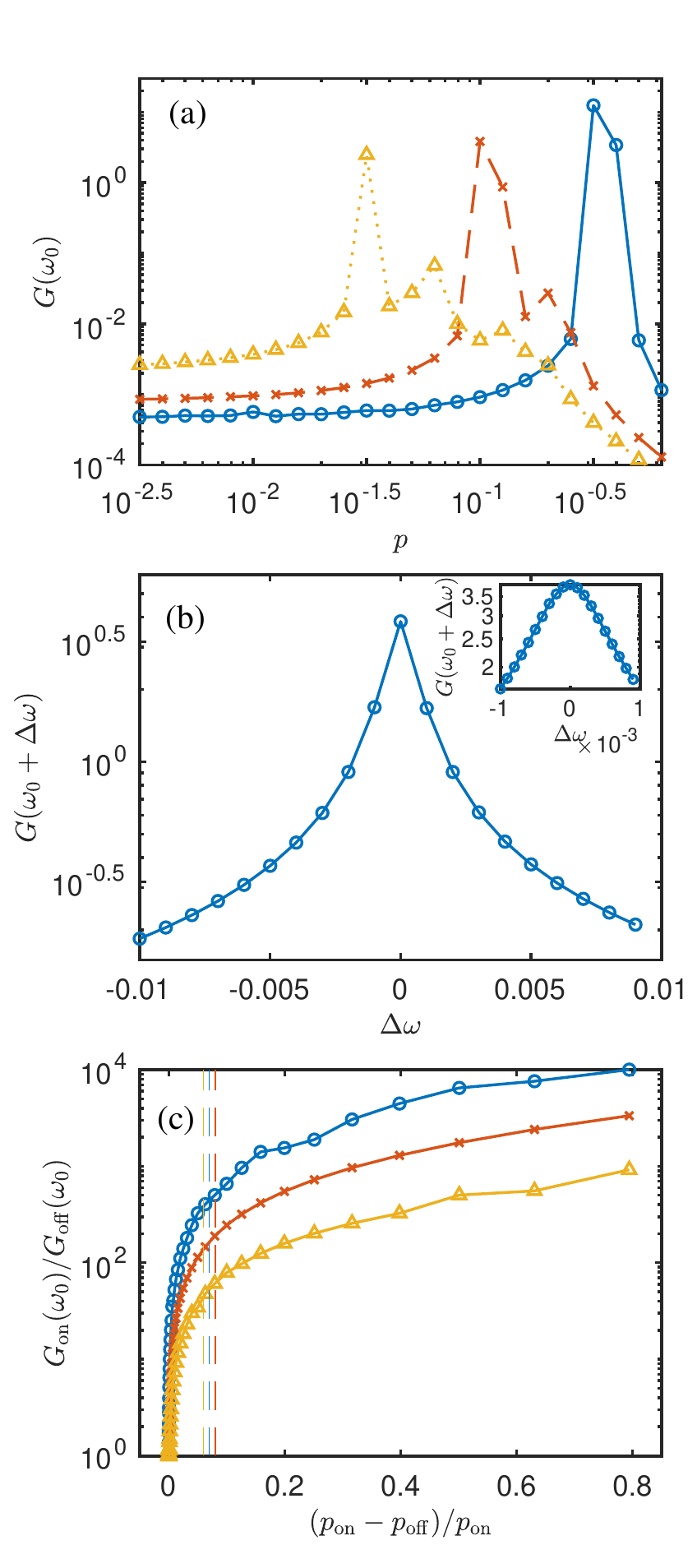}
\caption{(a) The gain $G(\omega_0)$  
for the acoustic device as a function of pressure $p$ for three values of the 
driving frequency $\omega_0=13.1$ (triangles), $14.9$ (exes), and $15.7$ (circles). (b) The gain $G(\omega_0+\Delta \omega)$ over a small frequency range 
$\Delta \omega$ near the driving frequency $\omega_0=14.9$. The inset is a close-up of the 
gain to within $10^{-3}$ of $\omega_0$. (c) The gain ratio $G_{\rm on}(\omega_0)/G_{\rm off}(\omega_0)$ as a 
function of the normalized change in pressure between the on and 
off states, $(p_{\rm on}-p_{\rm off})/p_{\rm on}$, for $p_{\rm on} = 
3.2\times10^{-1}$ (circles), $10^{-1}$ (exes), and $3.2 \times 10^{-2}$ 
(triangles) and the sizes of all particles are changed to control the 
pressure. The vertical dashed lines indicate the value of $(p_{\rm on}-p_{\rm off})/p_{\rm on}$ at which contacts would begin breaking if the size 
of only a single particle was changed to control the pressure. For all data, the 
driving amplitude 
is $A_0 = 10^{-6}$ and the damping parameter $b=10^{-3}$.}
\label{AllPart_GainvsP}
\end{figure}

\begin{figure}[h!]
\includegraphics[width=2.8in]{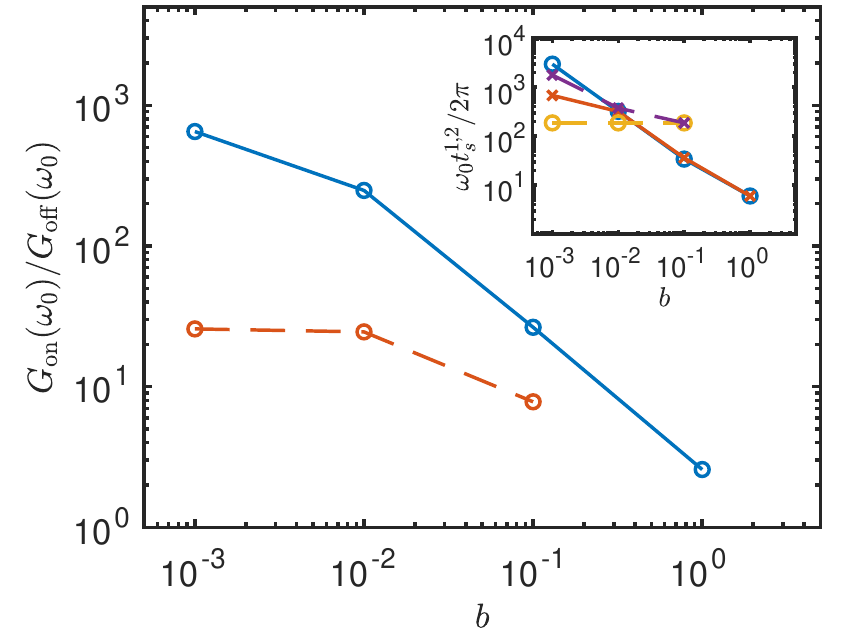}
\caption{The gain ratio $G_{\rm on}(\omega_0)/G_{\rm off}(\omega_0)$ 
between the on and off states 
versus the damping parameter $b$ at fixed driving frequency 
$\omega_0 = 14.9$ for pressure-induced switching (solid line) and $16.0$ 
for switching with contact breaking (dashed line). The inset 
shows the switching time $\omega_0 t_s^1/2\pi$ from the on to 
the off state (open circles) and $\omega_0 t_s^2/2\pi$ from the off to 
the on state (exes) versus 
$b$ for the same systems in (a).}
\label{AllPart_dGdt}
\end{figure}

\begin{figure}[h!]
\includegraphics[width=2.95in]{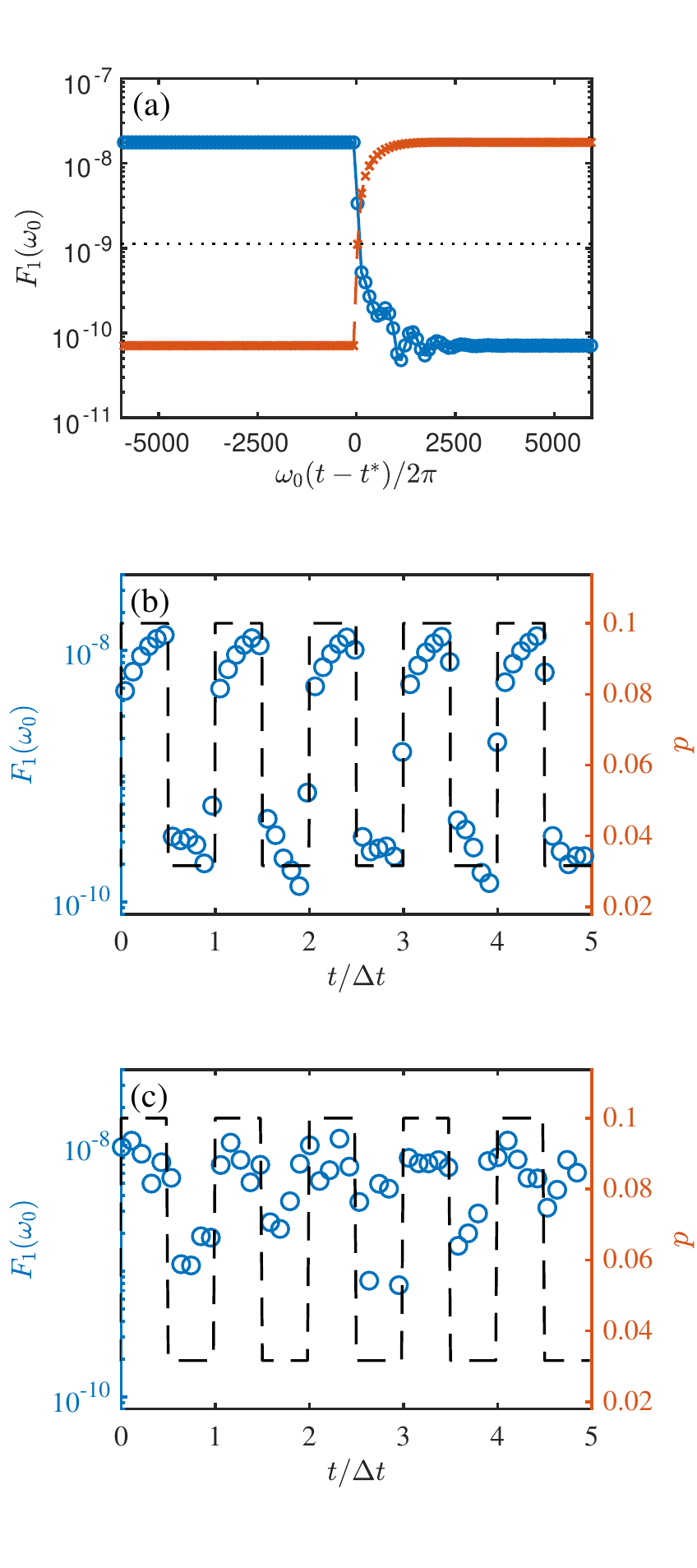}
\caption{(a) The Fourier transform $F_1(\omega_0)$ of the $x$-displacement of 
disk $1$ as a function of time $\omega_0(t-t^*)/2\pi$ when switching 
the device at time $t^*$ from the 
``on” (pressure $p=10^{-1}$) to “off” ($p=3.2 \times 10^{-2}$) states 
(circles) and vice versa (exes) using a damping 
coefficient $b=10^{-2}$. The horizontal dotted line indicates the geometric 
mean ${\overline F}_1(\omega_0)$ of the on and off values of $F_1(\omega_0)$. 
The switching times $t_s$ are obtained by finding when $F_1(\omega_0)$ 
crosses ${\overline F}_1(\omega_0)$. 
(b) The Fourier transform $F_1(\omega_0)$ (open circles and left axes 
labels) of the 
$x$-displacement of disk $1$ as a function of time $t/\Delta t$ (after 
reaching an initial steady state at $t=0$) during  
continuous 
switching of the device between the ``on” and “off” states using 
$b=10^{-2}$. The pressure of the device (dashed line and right axes labels) follows a square wave signal with $\Delta t/t_s \approx 3.7$. (c) Same as (b) except $\Delta t/t_s \approx 0.7$.}
\label{AllPart_switch}
\end{figure}

In Fig.~\ref{AllPart_Gain} (a), we show the Fourier transform
$F_1(\omega)$ of the $x$-displacement of the output disk $1$ in the
device after driving the input disk $3$ sinusoidally according to
Eq.~\ref{drive} with amplitude $A_0=10^{-6}$ and frequency
$\omega_0=14.9$.  Since displacing disk $3$ in the $x$-direction is
not a pure eigenmode of the mass-weighted dynamical matrix for the
full system, there are contributions to $F_1(\omega)$ over a wide
range of frequencies. Despite this, there is a strong response at the
driving frequency $\omega_0$.  We also show the Fourier transform
$F_3(\omega)$ of the $x$-displacement of the input disk $3$, and
calculate the gain $G(\omega_0)=F_1(\omega_0)/F_3(\omega_0)$.  We find
that the gain in this high pressure regime is independent of the
amplitude of the driving.  (See Fig.~\ref{AllPart_Gain} (b).) The gain
for the on state at high pressure $p=10^{-1}$ is $G(\omega_0) \approx
1$, whereas the gain for the off state at lower pressure $p=3.2\times
10^{-2}$ is more than two orders of magnitude smaller.  In
Fig.~\ref{AllPart_GainvsP} (a), we show the variation of the gain
$G(\omega_0)$ with pressure for several values of the driving
frequency $\omega_0$. We verify that we can accurately measure the
gain ($G(\omega_0) \approx 3.5$) near each resonance in
Fig.~\ref{AllPart_GainvsP} (b).  For each driving frequency, $\omega_0
= 13.1$, $14.9$, and $15.7$, the ratio of the maximum gain
(at $p_{\rm on}$, on state) and minimum gain (at $p_{\rm off}$, off
state) increases as a function of the normalized pressure difference
$(p_{\rm on}-p_{\rm off})/p_{\rm on}$.  For $\omega_0 =15.9$, the
increase in the gain ratio $G_{\rm on}(\omega_0)/G_{\rm
  off}(\omega_0)$ is the largest, reaching $10^4$ at the largest
pressure difference. (See Fig.~\ref{AllPart_GainvsP} (c).)  We can
also vary the gain ratio between the on and off states at fixed
driving frequency $\omega_0$ by changing the damping coefficient $b$.
In Fig.~\ref{AllPart_dGdt}, we show that the gain ratio decreases as a
power law with the damping parameter, $G_{\rm on}(\omega_0)/G_{\rm
  off}(\omega_0) \sim b^{-1}$ for large $b$.  In contrast, the gain
ratio plateaus in the limit of small $b$.

We have demonstrated that we can achieve gain ratios between the on
and off states for the acoustic device that are at least four orders
of magnitude. We will now analyze the ability of the device to switch
from the on to off states and vice versa. We will change the sizes of
all particles in the device to instantaneously increase or decrease
the pressure and induce switching. In Fig.~\ref{AllPart_switch} (a),
we show the Fourier transform $F_1(\omega_0)$ of the $x$-displacement
of disk $1$, while driving disk $3$ sinusoidally at $\omega_0$ in the
$x$-direction. We consider two situations: 1) The device is initiated
in the on state at pressure $p=10^{-1}$. The system remains in the on
state for a given amount of time. At time $t^*$, the system is
switched to the off state by decreasing the pressure to $3.2\times
10^{-2}$ and remains there. 2) The device is initiated in the off
state at $p=3.2\times 10^{-2}$ and remains in the off state for a
given amount of time.  At time $t^*$, the pressure is increased to
$p=10^{-1}$ and remains there.  In Fig.~\ref{AllPart_switch} (a), we
show that for the case of pressure-induced switching, the switching
time $t_s$ from on to off and from off to on are comparable. For
damping parameter $b = 10^{-2}$, $\omega_0 t_s/2\pi \approx 10^{3}$,
where $t_s$ is obtained by determining the time at which
$F_1(\omega_0)$ reaches the geometric mean of the values of
$F_1(\omega_0)$ in the on and off states.

Note that the switching time $t_s$ is rather large ($\sim 10^3$
oscillations for $b=10^{-2}$). This large timescale occurs because the
oscillation of a single input particle and a single output particle
are not pure eigenmodes of the mass-weighted dynamical matrix of the
device. Thus, when switching from the on to off state, there is
residual energy in the eigenmode at $\omega_0$ that must be removed
via damping. When switching from the off to on state, there is
residual energy in eigenmodes that are different from the one at
$\omega_0$ that must be removed via damping.  This picture is
consistent with the fact that the switching times scale as $\omega_0
t_s \sim b^{-1}$ as shown in the inset to Fig.~\ref{AllPart_dGdt}
(solid lines). With this scaling behavior, $t_s$ can be decreased by increasing
$b$. However, as shown in Fig.~\ref{AllPart_dGdt}, the gain ratio also
decreases with increasing $b$, which makes it difficult to distinguish
between the on and off states. Thus, the optimal performance for the
pressure-induced acoustic switch is the relatively small value for the
damping parameter, $b \approx 10^{-2}$, where the gain ratio no longer
increases dramatically with decreasing $b$, yet $t_s$ is relatively
small.

We also studied dynamic switching using a square wave input signal for
the time dependence of the pressure. In this case, the pressure is
large for given amount of time $\Delta t$ and then it is switched
instantaneously to a lower pressure for a time period $\Delta
t$. After an additional time period $\Delta t$, the pressure is again
switched back to the large pressure value. This process is then
repeated for a given number of cycles. (See the dashed lines in
Fig.~\ref{AllPart_switch} (b) and (c).) When $\Delta t$ satisfies
$\Delta t \gtrsim t_s$, $F_1(\omega_0)$ tracks with the pressure
signal and is nearly able to reach the steady-state values of
$F_1(\omega_0)$ at each pressure as shown in Fig.~\ref{AllPart_switch}
(b). (The steady-state values of $F_1(\omega_0)$ are $\approx 2\times
10^{-8}$ for the on state and $\approx 8\times 10^{-11}$ for the off
state.) For $\Delta t \lesssim t_s$, $F_1(\omega_0)$ is not able to
track the input signal (as shown in Fig.~\ref{AllPart_switch} (c)) and
thus the gain ratio between the on and off states for dynamic
switching is much smaller than the gain ratio in steady-state.  The
case $\Delta t \gg t_s$ is similar to the step function perturbation
in Fig.~\ref{AllPart_switch} (a).

\subsection{Switching with contact breaking}
\label{contact_breaking}

\begin{figure}[h!]
\includegraphics[width=3in]{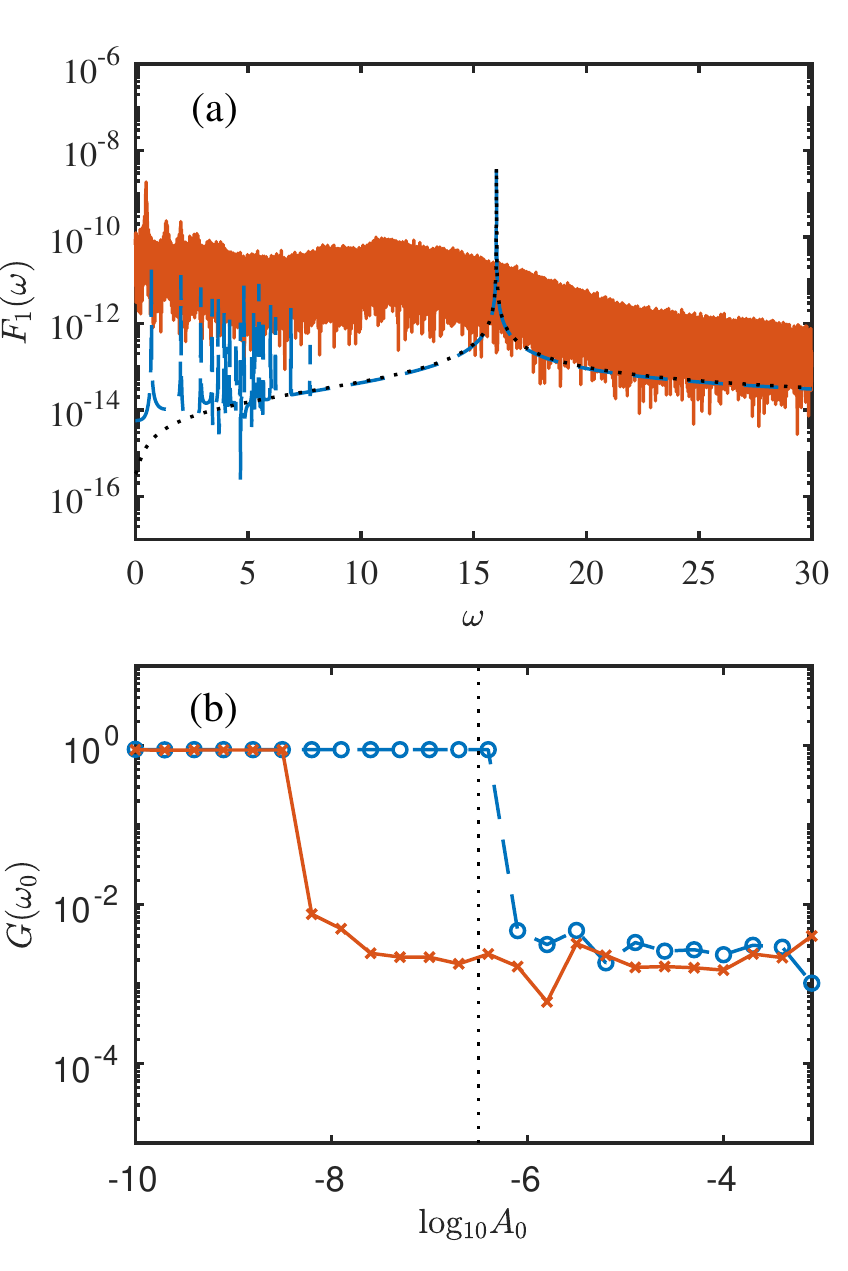}
\caption{
(a) The Fourier transform $F_1(\omega)$ of 
the $x$-displacement of disk $1$ for the device with pressure $p = 10^{-6}$ 
(dashed line) and $10^{-8}$ (solid line) obtained 
by driving disk $3$ sinusoidally with amplitude $A_0=3.2 \times10^{-7}$ 
and frequency $\omega_0=16.0$, using damping coefficient $b=10^{-3}$. The dotted line shows the Fourier transform 
$F_3(\omega)$ of the $x$-displacement of the input disk $3$. (b) The gain G($\omega_0$) (defined in Eq.~\ref{Gain}) plotted 
as a function 
of the driving amplitude $A_0$ with driving frequency $\omega_0=16.0$ 
for a device at pressure $p = 10^{-6}$ (solid line) and 
$10^{-8}$ (dashed line), using damping parameter $b=10^{-3}$. 
The vertical dotted line indicates the amplitude of 
the driving $A_0=3.2 \times10^{-7}$ in (a).}
\label{AllPart_Gain_CB}
\end{figure}

In this subsection, we describe the results for acoustic devices where
switching between the on and off states is achieved by changing the
network of interparticle contacts. In the systems we consider, the
interparticle contact network does not change during the vibrations in
the on state. However, the interparticle contact network fluctuates
during the vibrations in the off state. In Fig.~\ref{AllPart_Gain_CB}
(a), we show the Fourier transform $F_1(\omega)$ of the
$x$-displacement of disk $1$ for the device in the low-pressure regime
with $p=10^{-6}$ (on state) and $10^{-8}$ (off state) obtained by
driving disk $3$ sinusoidally with amplitude $A_0 = 3.2\times 10^{-7}$
and frequency $\omega_0 = 16.0$, using damping coefficient
$b=10^{-3}$.  $F_1(\omega)$ for the device at $p=10^{-6}$ is similar
to that in the high pressure regime (Fig.~\ref{AllPart_Gain} (a)).
However, $F_1(\omega)$ at $p=10^{-8}$ has a broad and noisy spectrum
since the interparticle contact network fluctuates during the
vibrations. (See the contact-breaking regime for $D(\omega)$ in
Fig.~\ref{vCorr} (b).)  In the low-pressure regime, the device can be
switched on and off by varying the amplitude of the driving at fixed
frequency $\omega_0$.  In Fig.~\ref{AllPart_Gain_CB} (b), we show the
gain $G(\omega_0)$ of the device at pressures $p=10^{-6}$ and
$10^{-8}$ and driving frequency $\omega_0 = 16.0$. At small driving
amplitudes, the gain is relatively large, $G(\omega_0) \approx 1$.  As
the amplitude is increased, changes in the interparticle contact
network begin to occur at a characteristic amplitude $A_0^*$ that
scales with pressure. See Fig.~\ref{ThreePort} (b) for a device in
which the interparticle contact network has fewer contacts in the off
state than in the on state. For example, $A_0^* \approx 10^{-8.5}$ for
$p=10^{-8}$ and $A_0^* \approx 10^{-6.5}$ for $p=10^{-6}$.  The onset
of contact breaking causes the gain to drop abruptly by more than two orders of
magnitude.  We show in Fig.~\ref{AllPart_Gain_CB} (b) that if we drive
the device at amplitude $A_0 = 3.2\times 10^{-7}$ and frequency
$\omega_0$, it is in the on state at pressure $p=10^{-6}$ and the off
state at $10^{-8}$. We can obtain similar behavior if we drive the
device in the amplitude range $5 \times 10^{-9} \lesssim A_0 < 3.2
\times 10^{-7}$.

\begin{figure}[h!]
\includegraphics[width=2.95in]{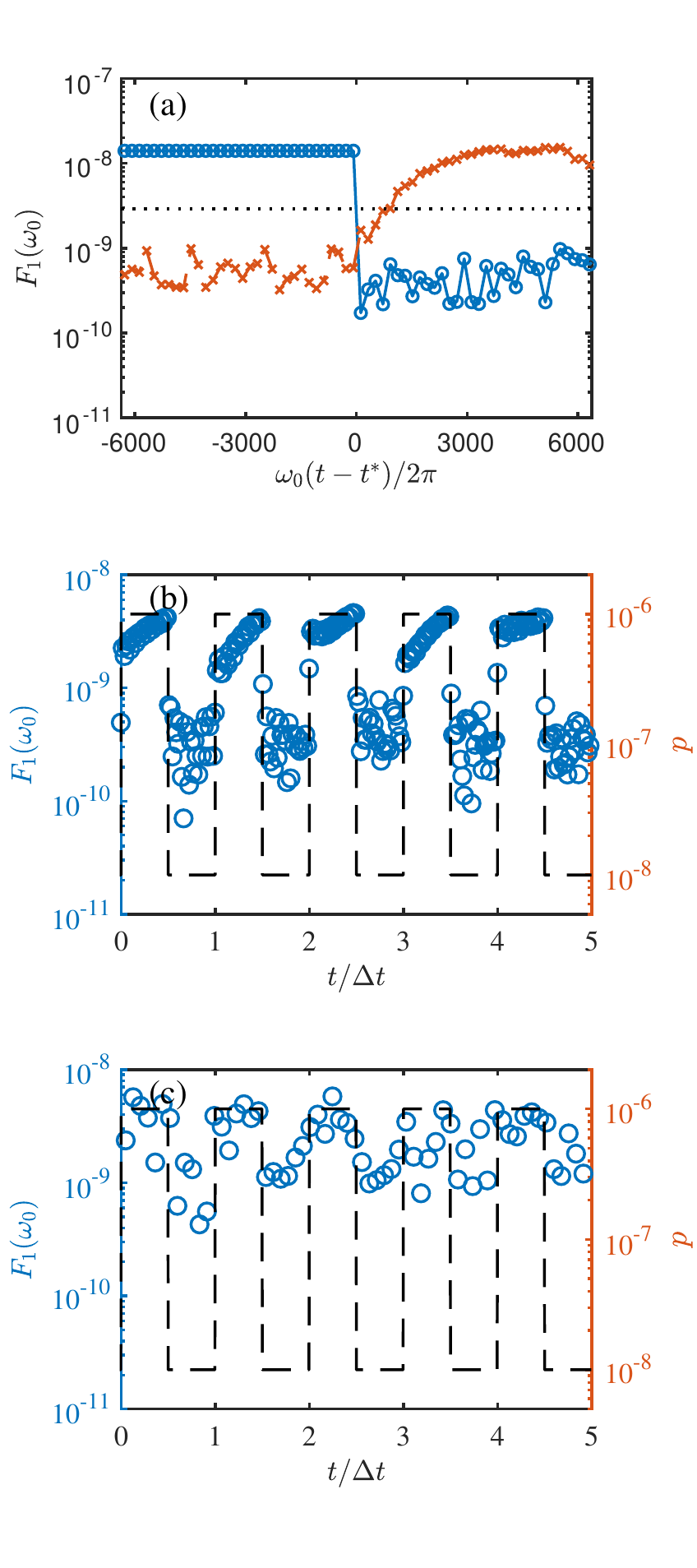}
\caption{(a) The Fourier transform $F_1(\omega_0)$ of the $x$-displacement of 
disk $1$ as a function of time $\omega_0 (t-t^*)/2\pi$ when switching the device from the 
``on” (pressure $p=10^{-6}$) to “off” ($p=10^{-8}$) states 
(circles) and vice versa (exes) at time $t^*$ using a damping 
coefficient $b=10^{-3}$. The driving frequency and amplitude are $\omega_0 
=16.0$ and $A_0 = 3.2 \times 10^{-7}$. (b) The Fourier transform 
$F_1(\omega_0)$ of the 
$x$-displacement of disk $1$ (circles and left axes labels) as a function of time $t/\Delta t$ when 
continuously switching the device between the ``on” and “off” states 
using 
$b=10^{-3}$. The dashed line shows the pressure of the device (right 
axes labels), which has 
a square wave form with $\Delta t/t_s \approx 2.7$ (where $t_s$ is 
the time for the device to switch from the off to the on states). (c) Same 
as (b) except $\Delta t/t_s \approx 0.68$. For (a)-(c), the off and 
on states possess different interparticle contact networks.}
\label{AllPart_switch_CB_1}
\end{figure}

We show the ratio of the gain in the on versus the off state $G_{\rm
  on}(\omega_0)/G_{\rm off}(\omega_0)$ as a function of the damping
parameter $b$ for devices that experience contact breaking in
Fig.~\ref{AllPart_dGdt}.  As for devices with no contact breaking, the
gain ratio decreases with $b$ for large $b$, whereas it forms a
plateau for small $b$. However, at small $b$, the gain ratio is nearly
two orders of magnitude smaller for devices that incorporate contact
breaking compared to those that do not.

In Fig.~\ref{AllPart_switch_CB_1} (a), we show the performance of the
acoustic device in switching from the on to off states and vice versa
using damping parameter $b=10^{-3}$ in the regime where contact
breaking occurs.  An interesting feature is that the times $t^1_s$ and
$t^2_s$ for switching the device from the on to the off state
and from the off to the on state, respectively, are different.  As shown in the
inset in Fig.~\ref{AllPart_dGdt}, the switching time from the on to
the off state, $\omega_0 t^1_s/2\pi \sim 10^2$, is nearly independent of
the damping parameter $b$, and is less than the switching time from
the off to the on state ($t^1_s < t^2_s$) since $t^2_s$ grows with
decreasing $b$.

We show the results for dynamic switching with contact breaking for
the device in Fig.~\ref{AllPart_switch_CB_1} (b) for the case $\Delta
t/t^1_s \approx 2.7$. $F_1(\omega_0)$ can roughly track the pressure
signal, but the signal for the off state is noisy. When we decrease
$\Delta t$ such that $\Delta t/t^1_s \approx 0.68$, there is no
significant difference between $F_1(\omega_0)$ in the on and off
states and $F_1(\omega_0)$ is not strongly affected by the relatively 
rapid changes in 
pressure.

\begin{figure}[h!]
\includegraphics[width=3in]{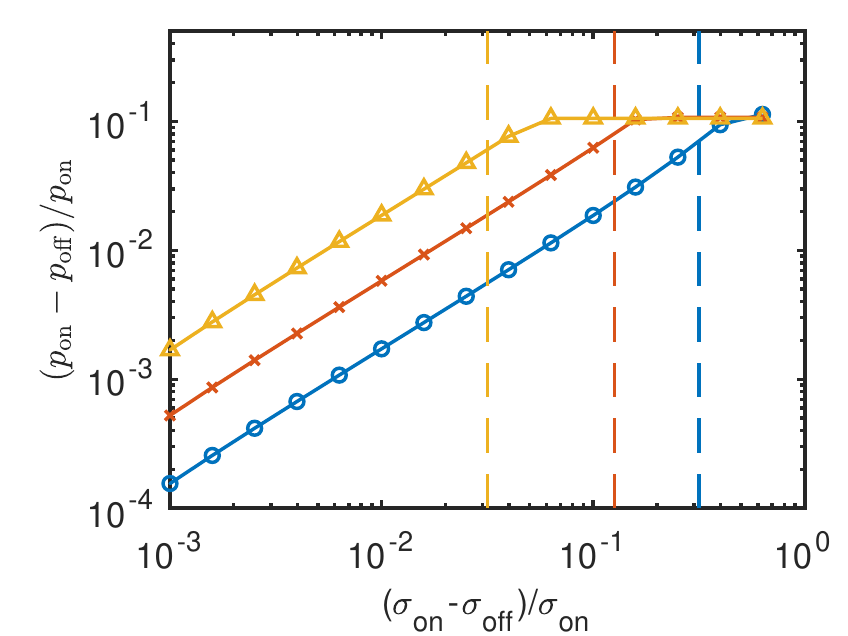}
\caption{The normalized change of pressure 
$(p_{\rm on} - p_{\rm off})/p_{\rm on}$, 
where $p_{\rm on}$ and 
$p_{\rm off}$ are the pressures in the on and off states, respectively, as a
function of the normalized change in the size $(\sigma_{\rm on} - \sigma_{\rm off})/
\sigma_{\rm on}$ of a single 
control particle for pressures $p_{\rm on} =
10^{-2}$ (circles), $3.2 \times 10^{-2}$ (exes), and
$10^{-1}$ (triangles). The vertical dashed lines (from left to right) 
indicate the change in 
size above which the control particle loses a contact with neighboring 
particles for $p_{\rm on}=10^{-2}$, $3.2 \times 10^{-2}$, and $10^{-1}$.}
\label{dpdr_3Part}
\end{figure}

\subsection{Single-particle control signal}
\label{single_particle}

For systems without contact breaking, the gain ratio between the on
and off states $G_{\rm on}(\omega_0)/G_{\rm off}(\omega_0)$ is
determined by the difference in pressure that can be achieved, for
example, by changing all particle sizes.  In
Fig.~\ref{AllPart_GainvsP} (c), we showed that the gain ratio
increases with the normalized pressure difference $(p_{\rm on}-p_{\rm
  off})/p_{\rm on}$.  When we use all particles in the device to
change the pressure, we can achieve a wide range of normalized
pressure differences from $0$ to $0.8$, and thus we can obtain a wide
range of gain ratios from $1$ to $10^4$. However, when we use only a single
control particle (e.g. the central grain in Fig.~\ref{ThreePort} (b)), the maximum change in the
normalized pressure that can be achieved scales as $1/N$.  In
Fig.~\ref{dpdr_3Part}, we show that for $N=30$, the maximum normalized
pressure difference is $\sim 10^{-1}$ using a single control
particle. Operating the device in the regime where the interparticle
contact network remains intact further restricts the normalized
pressure difference that can be used.  If we limit $(p_{\rm on}-
p_{\rm off})/p_{\rm on} < 10^{-1}$, the maximum gain ratio that can be
achieved is $G_{\rm on}(\omega_0)/G_{\rm off}(\omega_0) \approx
10^{2.5}$, which is less than the value of $10^{4}$ achieved for
devices that change the sizes of all particles. (See the vertical lines in 
Fig.~\ref{AllPart_GainvsP} (c).)

\begin{figure}[h!]
\includegraphics[width=3in]{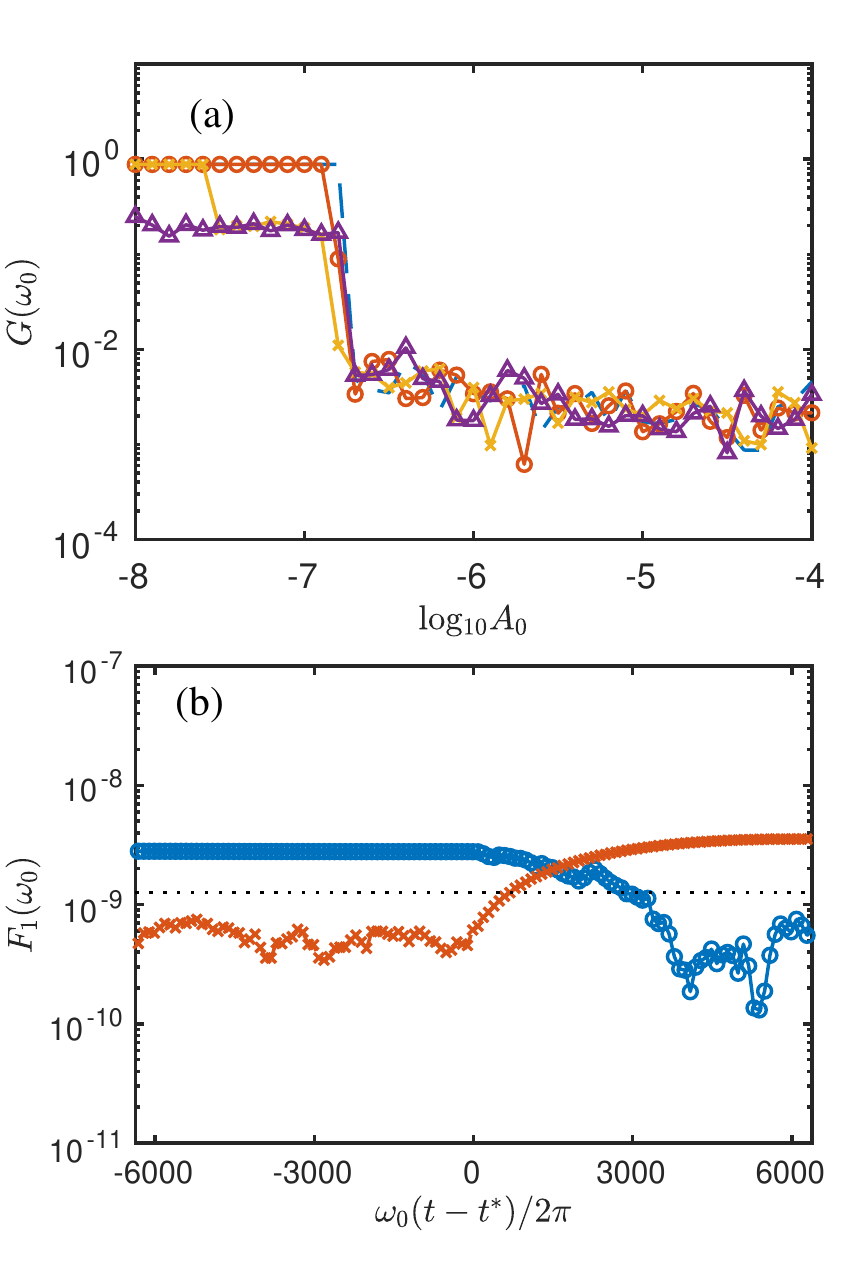}
\caption{
(a) The gain G($\omega_0$) (defined in Eq.~\ref{Gain}) plotted 
as a function 
of the driving amplitude $A_0$ at fixed driving frequency $\omega_0=16.0$ 
for a device with a single control particle at $\Delta \sigma/\sigma = 0$ 
(dashed line),  $5.1 \times 10^{-7}$ (circles), $5.4 \times 10^{-7}$ (exes), 
and $5.5 \times 10^{-7}$ (triangles), using damping parameter $b=10^{-3}$.
(b) The Fourier transform $F_1(\omega_0)$ of the $x$-displacement of 
disk $1$ as a function of time $\omega_0 (t-t^*)/2\pi$ when switching 
the device from the 
on ($\Delta \sigma/\sigma=0$) to off 
($\Delta \sigma/\sigma=5.4 \times10^{-7}$) state 
(circles) and vice versa (exes) at time $t^*$ using a single control 
particle and damping parameter $b=10^{-3}$. The driving frequency and amplitude are $\omega_0 
=16.0$ and $A_0 = 6.3 \times 10^{-8}$, respectively.}
\label{3Part_GainA0_CB}
\end{figure}
 
As expected, the performance of devices that only have a single
control particle is also degraded in the regime where contact breaking
occurs.  In Fig.~\ref{3Part_GainA0_CB} (a), we show the gain
$G(\omega_0)$ for a device driven at frequency $\omega_0 = 16.0$
versus the amplitude $A_0$ and compare it to the gain from systems in
which the size of a single control particle has been decreased by an
amount $\Delta \sigma/\sigma$.  The reference system (with $\Delta
\sigma/\sigma=0$) is in the on state with $G(\omega_0) \approx 1$ for
small driving amplitudes. As the driving amplitude increases, the gain
decreases abruptly when the interparticle contact network begins to
fluctuate. Similar behavior is found in Fig.~\ref{AllPart_Gain_CB}
(b). When the change in the size of the control particle is small,
i.e. $\Delta \sigma/\sigma = 5.1 \times 10^{-7}$, $G(\omega_0)$ is
similar to that for the reference system. When $\Delta \sigma/\sigma$
is increased further, $G(\omega_0)$ develops an intermediate plateau
between that for the on state ($G(\omega_0) \approx 1$) and the off
state ($G(\omega_0) \approx 10^{-3}$).  Thus, by changing the size of
only one particle, a dynamic state with an intermediate value of the
gain occurs. This intermediate state represents a system in which only
the contacts that involve the control particle (not all interparticle
contacts) are fluctuating.  As shown in Fig.~\ref{3Part_GainA0_CB}
(b), the presence of the state with intermediate gain significantly
reduces the difference in $F_1(\omega_0)$ between the on and off
states during switching.  For all changes in pressure that can be
achieved using a single control particle and induce contact breaking
between the control particle and its neighbors, we find a dynamical
state with intermediate gain between that for the on state (with no
contact breaking) and off state (with contact breaking among all
particles).

\begin{figure}[h!]
\includegraphics[width=3in]{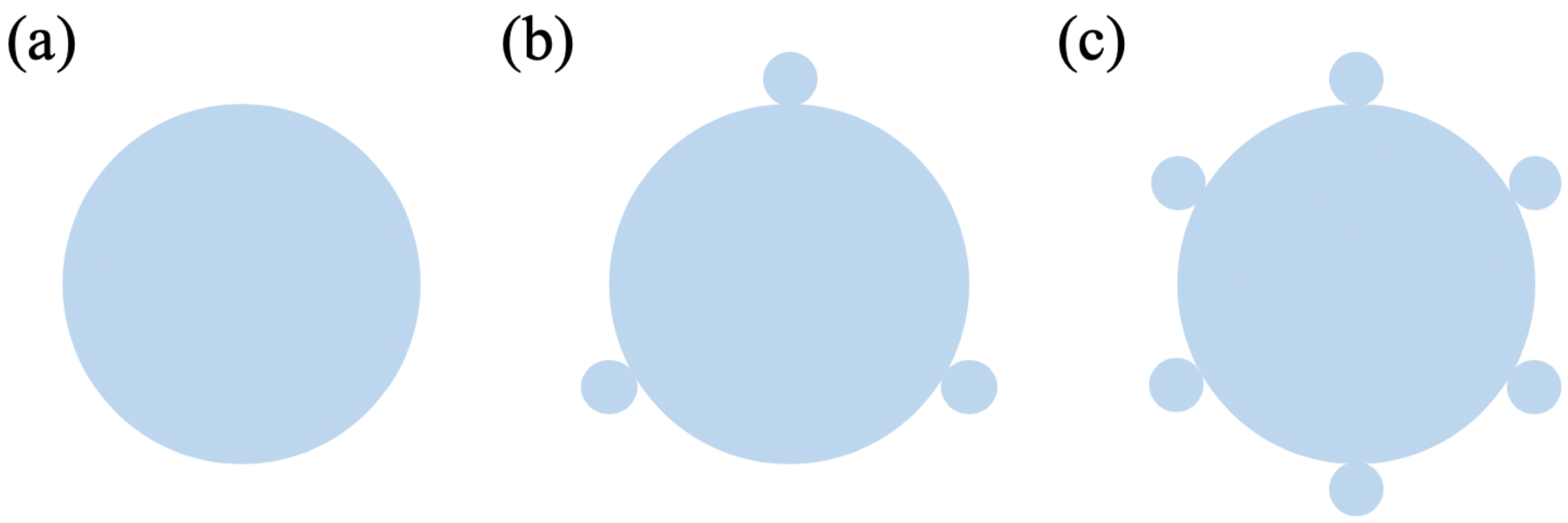}
\caption{Circular particles with different numbers of circular 
knobs, (a) $n=0$, 
(b) $3$, and $6$, placed symmetrically around the perimeter of the 
particle. The central disk has diameter $\sigma$ and the circular knobs 
have diameter $\sigma_k =
(2\sqrt{3}/3-1)\sigma$.}
\label{Knobs}
\end{figure}

\subsection{Generating 2D granular crystals}
\label{periodic}

In Sec.~\ref{pressure_switching}, we described the performance of
acoustic switching devices composed of $N=30$ monodisperse disks (with
diameter $\sigma$) of two different masses ($N_L= 21$ with mass $m_L$
and $N_S=9$ with mass $m_S$) arranged on a two-dimensional hexagonal
lattice similar to that in the inset of Fig.~\ref{BandGap}. To realize
these devices in experiments, an automated method of making the 2D
granular crystals must be developed. Methods for generating granular
crystals in experiments include vibration~\cite{reis}, cyclic
shear~\cite{kudrolli}, and combinations of vibration and
shear~\cite{daniels}. However, it is well-known that generating
defect-free granular crystals is difficult, requiring an exponentially
large number of small amplitude vertical vibrations or shear
cycles~\cite{sid}. Further, one way to generate a large frequency band
gap in granular crystals is to choose grains with large mass
ratios. However, vibration and shear in systems composed of grains
with large mass ratios often give rise to de-mixing or segregation,
where grains with similar masses cluster together~\cite{gray,hill},
instead of forming the alternating pattern of grains with large and
small masses shown in the inset of Fig.~\ref{BandGap} that maximizes
the width of the frequency band gap.

\begin{figure}[h!]
\includegraphics[width=3in]{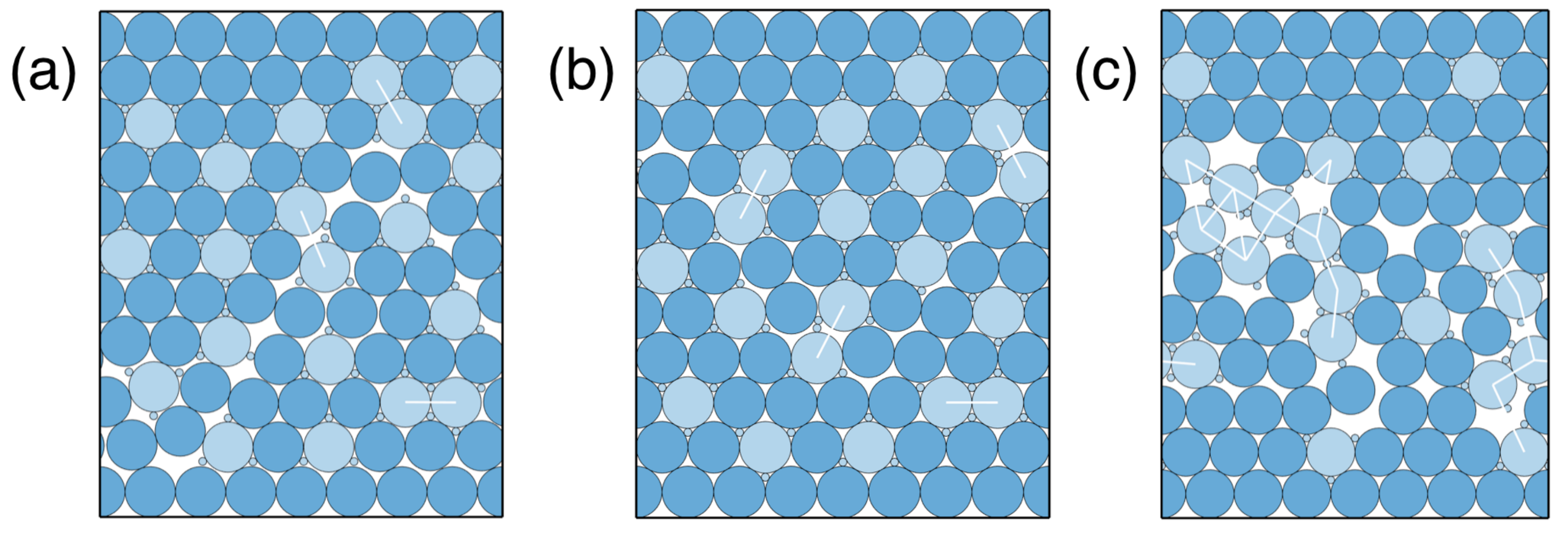}
\caption{Snapshots of disk configurations from the cooling simulations 
that contain disks without (dark) and with 
$n=3$ knobs (light). The configurations possess (a) $N_{kk} = 3$, (b) $4$, and $17$ 
contacts between the grains with knobs, which are indicated by 
solid white lines.}
\label{Knobs_config}
\end{figure}

In this subsection, we describe a method to enhance crystallization
into the alternating pattern of grains shown in the inset to
Fig.~\ref{BandGap}, and implement it in numerical simulations. We
consider mixtures of $N_L=21$ disks with diameter $\sigma$ and $N_S=9$
disks with the same size and mass of the others, but they possess
small circular knobs symmetrically placed around their perimeter. (See
Fig.~\ref{Knobs}.) The knobs have diameter $\sigma_k =
(2\sqrt{3}/3-1)\sigma$ and the angular separation between the knobs is
$2\pi/n$, where $n=0$, $3$, and $6$ gives the number of knobs. The
size and spacing of the knobs is chosen so that they fit within the
instertices of the circular grains without knobs arranged on a
hexagonal lattice. The knobs will only fit within the interstices when
they are surrounded by grains without knobs.  (See
Fig.~\ref{Knobs_config}.) Thus, in mixtures of grains with and without
knobs, there is an effective repulsion between grains with knobs that
enhances crystallization into the alternating pattern in the inset of
Fig.~\ref{BandGap}.

As we will show below, we are able to create packings in which grains
with knobs and grains without knobs form an altenating pattern on a
hexagonal lattice.  If, in experiments, the grains with knobs are made
of a composite material for which part of the material can be
preferentially dissolved, the knobs, as well as part of the core of
the grain, can be dissolved after the hexagonal assembly has been
generated. Thus, this procedure can generate an alternating pattern of 
light and heavy grains. 

\begin{figure}[h!]
\includegraphics[width=3in]{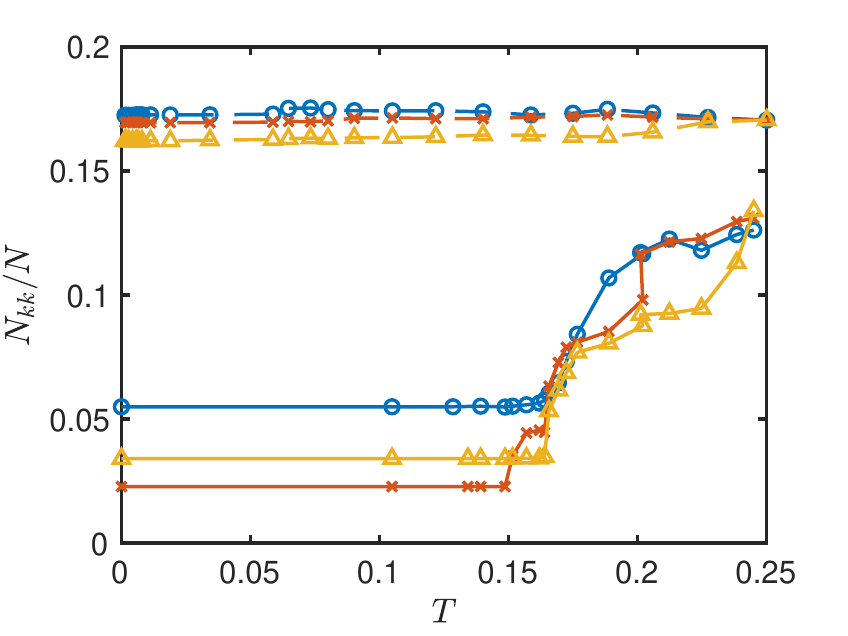}
\caption{Fraction of contacts between grains with knobs $N_{kk}/N$ as a 
function of temperature $T$ during cooling with damping parameter $b=0.05$ in conventional 
MD simulations (dashed lines) and during cooling with $b=0.05$ using umbrella sampling 
(solid lines) in systems with $n = 0$ (circles), $3$ (exes), and $6$ 
(triangles) knobs. The curves are averaged over $50$ initial conditions.}
\label{NkkvsT}
\end{figure}

\begin{figure}[h!]
\includegraphics[width=3in]{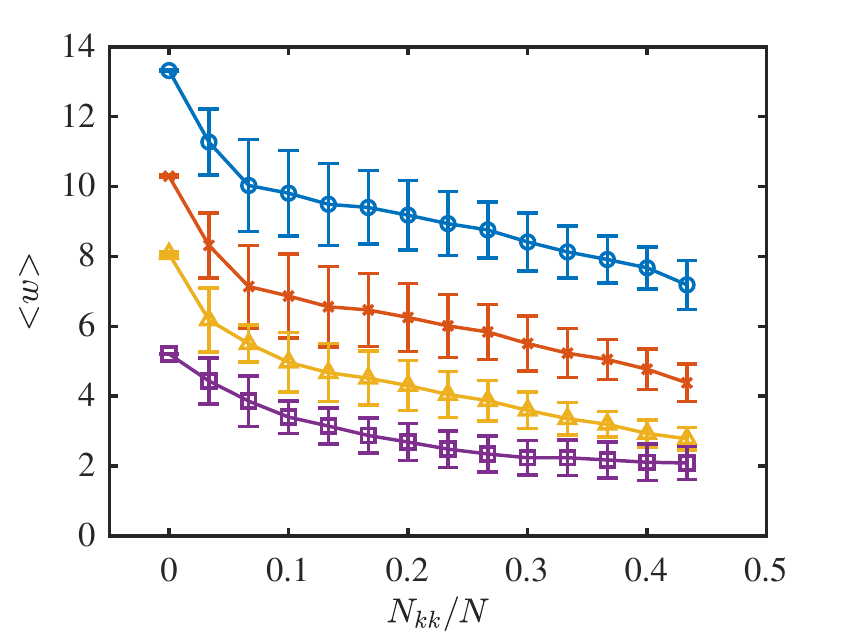}
\caption{The average maximum difference between adjacent eigenfrequencies $\langle w\rangle$ of the 
mass-weighted dynamical matrix versus the number of contacts 
between the grains with knobs $N_{kk}/N$ for a hexagonal packing with $N_L=21$ and $N_S=9$ and 
mass ratio $m_L/m_S = 100$ (circles), $20$ (exes), $10$ 
(triangles), and $5$ (squares). The means and standard deviations (error 
bars) are obtained by averaging over $50$ configurations in which the masses of the grains are chosen 
randomly as either $m_L$ and $m_S$ to yield a given $N_{kk}/N$.}
\label{band_Knobs}
\end{figure}

To measure the degree to which a disk configuration in the simulations
matches the alternating pattern in the inset to Fig.~\ref{BandGap}, we
determine the number of contacting pairs of grains with knobs, $N_{kk}$.
(See Fig.~\ref{Knobs_config} for configurations with different values
of $N_{kk}$.) A contact between grains with knobs means that the Voronoi
polygons of the grains share an edge, where the Voronoi tessellation
is based on the particle centers of the grains with and without 
knobs~\cite{oger}. The alternating pattern in the inset to
Fig.~\ref{BandGap} has $N_{kk}=0$, and $N_{kk} >0$ for configurations
with significant differences with the alternating pattern of grains 
with and wihtout knobs. Note that there are some configurations with
$N_{kk}=0$ that do not perfectly match the alternating pattern in the
inset to Fig.~\ref{BandGap}. However, we show in Fig.~\ref{band_Knobs}
that the average width of the bandgap $\langle w \rangle$ is
well-defined when we average over an ensemble of configurations with
the same $N_{kk}$.

As previously shown in Fig.~\ref{BandGap}, $w$ increases with the mass
ratio $m_L/m_S$. In addition, we find that $\langle w\rangle$
increases as $N_{kk} \rightarrow 0$, reaching a maximum that depends
on the mass ratio. Thus, especially for small mass ratios, it is
necessary to have packings with $N_{kk} \rightarrow 0$ to achieve
robust band gaps. In Appendix B, we discuss the performance of
acoustic switching devices (made from 2D granular crystals) with small
band gaps.

To generate packings of grains with and without knobs, we perform
discrete element simulations at constant pressure beginning at high
temperature in the liquid state. We then cool the system to low
temperature as a function of the cooling rate, which we adjust by
varying the damping parameter $b$.  The interactions between the large
circular disks, between the large disks and small knobs, and between
the small knobs have the same form as Eq.~\ref{Ep}.  Even though we
varied the cooling rate over several orders of magnitude, we did not
find a signficant decrease in $N_{kk}/N$ from its value in the liquid
state as shown in Fig.~\ref{NkkvsT} for $n=0$, $3$, and $6$.

To obtain configurations with small $N_{kk}/N$, we implemented an
umbrella sampling method to enhance the probability of rare
events~\cite{souaille,gonzalez,mezei}. We started with
$i=1,\ldots,N_c$ independent configurations at high temperature with
$N_{kk}/N \approx 0.16$. We then evolved each of these configurations using
the constant pressure cooling simulations until the $N_{kk}/N$ during
one of the trajectories $i=i^*$ decreases from the starting value by a fixed amount $\Delta n = \Delta
N_{kk}/N=0.01$.  We no longer consider the other $N_c-1$ trajectories and
instead spawn $N_c-1$ new ones using the same particle coordinates,
but each with different sets of particle velocities chosen randomly
from a Gaussian distribution with the temperature set to that when the
trajectory $i^*$ was identified. This process is then repeated
$1/\Delta n$ times. We show $N_{kk}/N$ versus temperature for the 
umbrella sampling method in Fig.~\ref{NkkvsT} for $n=0$, $3$, and $6$. 
We find that having three knobs allows the system to reach smaller values 
of $N_{kk}$ than having six knobs.  

\section{Conclusions and Future Directions}
\label{conclusions}

In this article, we describe active acoustic transistor-like devices
that can switch from the on to off states or vice versa using 2D
granular crystals.  We focus on systems composed of two types of
grains with the same size but different masses, since they possess
frequency band gaps in the vibrational density states that can be tuned by
the mass ratio $m_L/m_S$ and arrangement of heavy and light grains.
The input signal is generated by oscillating a grain at one side of
the device and measuring the resulting output signal from a grain on the
other side of the device.  The device can be switched between the on and
off states by changes in the size of one or many grains, which
controls the pressure. Switching can be achieved through two
mechanisms: 1) pressure-induced switching in which the on and off
states have the same interparticle contact networks and 2) switching
with contact breaking, where the interparticle contact networks are
different in the on and off states. In general, we find that the
performance of pressure-induced switching is better, with larger gain
ratios between the on and off states, than those for switching with
contact breaking. However, there is a tradeoff between large gain
ratios and fast switching times. Large gain ratios occur at small
damping parameters and fast switching times occur at larger damping
parameters. Even so, for pressure-induced switching, 2D granular
crystals can achieve gain ratios greater than $10^4$, and switching
times $\omega_0 t_s$ that represent $10^3$ oscillations at the driving
frequency. This switching time is comparable to that obtained
recently for sonic crystals~\cite{alagoz} and less than that for photonic
transistor devices~\cite{huang}.

Granular crystals are difficult to make in an automated way in
experiments.  We thus developed techniques to improve the efficiency
of making hexagonal crystals with an alternating pattern of heavy and
light grains.  The first improvement involved studying mixtures of
grains with and without small knobs arranged on their perimeter. The
size and arrangement of the knobs are chosen so that they fit in the
intertices between contacting grains without knobs. Since the grains
with knobs do not pack efficiently when they are next to each other,
there is an effective repulsion between the grains with knobs. The
similarity between a given configuration and the optimal configuration
with an alternating pattern can be measured using the fraction of
contacts between grains with knobs, $N_{kk}/N$. Using conventional
discrete element simulations of these mixtures undergoing cooling at
fixed pressure, $N_{kk}$ does not decrease significantly with
temperature.  However, when we apply an umbrella sampling-like
technique, we find that we can achieve $N_{kk}/N \rightarrow
0$. Further, we show that grains with $n=3$ knobs leads to smaller
values of $N_{kk}$ than that with $n=6$ when cooling with the improved
sampling technique.  Thus, our results encourage experimental studies of
mixtures of grains with and without knobs undergoing vertical
vibration or cyclic shear to study crystallization into hexagonal
crystals. After generating the alternating pattern of grains with and 
without knobs, the core regions of the grains with knobs and the knobs 
themselves can be dissolved away, yielding $m_L/m_S >1$. Experiments can
then be performed to measure the vibrational density of states in
these crystalline granular assemblies.

There are a number of important directions that we will pursue in
future studies. First, we will consider 3D granular crystals, which
have a broader range of mechanically stable crystal structures with
different symmetries, packing fractions, and numbers of nearest
neighbors. For example, we will determine the performance of FCC, BCC,
and HCP crystals with different mass distributions. Second, in the
current study, both the input and output signals oscillated in the
$x$-direction. In future studies in 3D, we can consider an input
signal that oscillates in a different direction than the measured
output signal.  A key aspect of these studies will be to understand the
spatial structure of the eigenmodes of the mass-weighted Hessian of
the device, and their overlap with the input and output
signals. Third, in the current modeling studies, we neglected
static friction. However, granular crystals in experiments have finite
friction, and thus it is important to understand how static friction
and the coupling of particle rotation and translation affect the switching
performance of the device. Fourth, an interesting application is to
create logical circuits from coupled acoustic switches that connect the
output of one device to the input of another. In future studies, we
will develop numerical implementations of coupled 2D granular crystals
that can perform logical operations.

\begin{figure}[h!]
\includegraphics[width=3in]{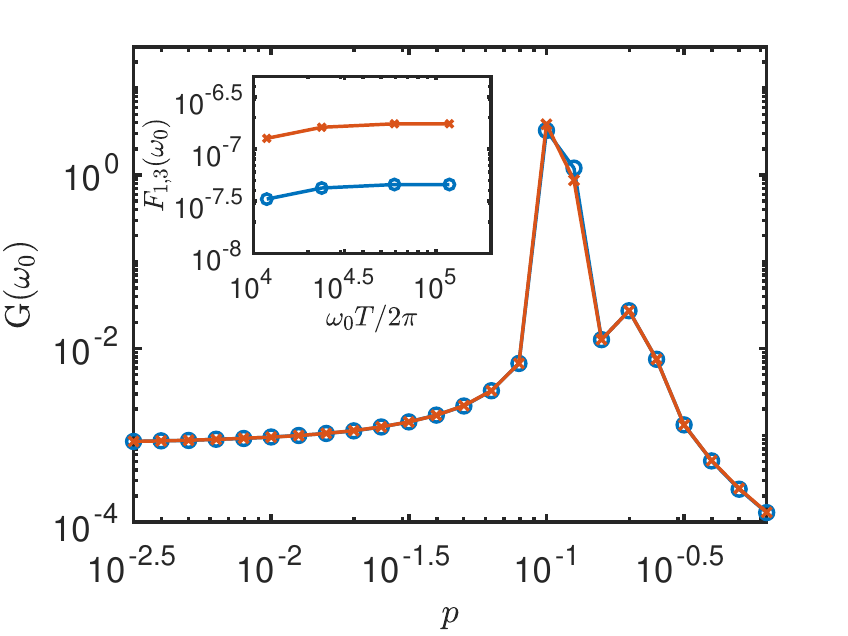}
\caption{The gain $G(\omega_0)$ for the acoustic device as a function
of pressure $p$ (for a system with no contact breaking) using a
total simulation time $\omega_0 T/2\pi$ = $10^{4}$
(circles) and $10^{5}$ (squares). The inset shows the Fourier transforms 
of the output and input signals,
$F_1(\omega_0)$ (exes) and $F_3(\omega_0)$ (circles), as a function of 
$\omega_0 T/2\pi$ for the device with pressure
$p=10^{-1}$. For all data, $\omega_0=14.9$, $A_0 = 10^{-6}$,
$b=10^{-3}$, and the sampling time
$\omega_0\Delta/2\pi$=$5.9\times10^{-3}$.}
\label{FFT_TimeDepend}
\end{figure}

\section*{Appendix A: Robustness of the Measurement of the Fourier Transforms 
of the Input and Output Signals}

Many of the results reported in this article depend on the
accurate calculation of the Fourier transform of the input and output signals
from particles $1$ and $3$, $x_{1,3}(t)-x_{1,3}^0$, respectively, where $x_{1,3}^0$ is 
the $x$-position
of particles $1$ and $3$ in the initial mechanically stable packing. We
calculate the Fourier $F_{1,3}(\omega) = \int_0^{\infty} [x_{1,3}(t)-x_{1,3}^0]
e^{i\omega t} dt$ numerically via the discrete Fourier transform:
\begin{equation}
F_{1,3}(\omega(l)) =\sum_{n=0}^{M-1} [x_{1,3}(n\Delta)-x_{1,3}(0)] e^{-i\cdot 2\pi l n/M},
\end{equation} 
where $\omega(l) = 2\pi l/T$, $M=T/\Delta$, $l$, and $n$ are
integers, $T$ is the total time of the input/output signals, and $\Delta$ is
the time interval between samples. In this Appendix, we calculate the gain $G(\omega_0)$ as a
function of the total time $T$ and sampling time $\Delta$ of the input
and output signals to show that our calculations do not depend
strongly on these parameters.

\begin{figure}[h!]
\includegraphics[width=3in]{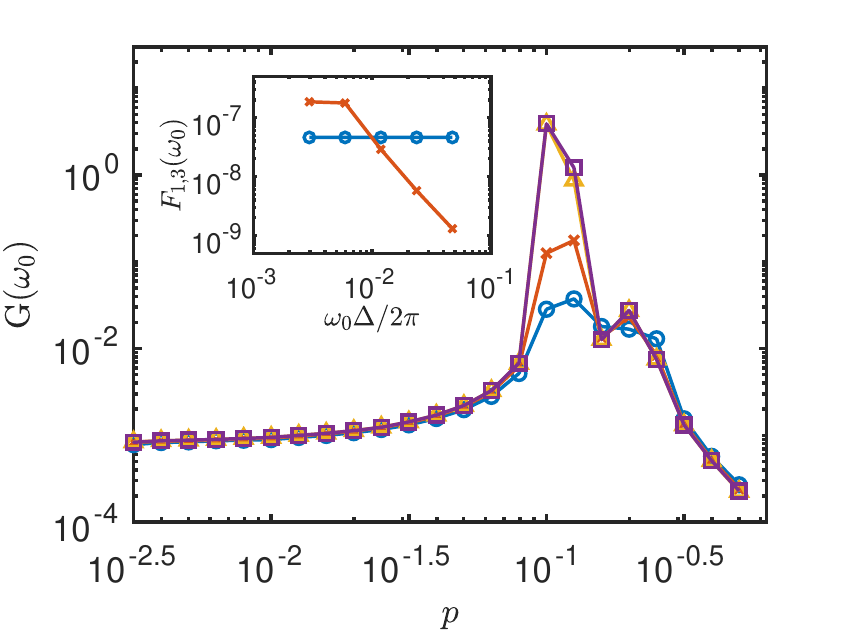}
\caption{The gain $G(\omega_0)$ for the acoustic device as a function of
pressure $p$ (for a system with no contact breaking) measured with sampling interval $\omega_0
\Delta/2\pi$ = $4.7 \times 10^{-2}$ (circles), $2.4 \times
10^{-2}$ (exes), $5.8 \times 10^{-3}$ 
(triangles), and $2.9 \times 10^{-3}$ (squares). The inset
shows the Fourier transforms of the output and input signals, $F_1(\omega_0)$ 
(exes) and $F_3(\omega_0)$ (circles), as a function of $\omega_0 \Delta/2\pi$ 
for the device with pressure
$p=10^{-1}$. For all data, $\omega_0=14.9$, $A_0 = 10^{-6}$,
$b=10^{-3}$, and the total simulation time $\omega_0
T/2\pi$=$5.9\times10^{4}$.}
\label{FFT_DeltaDepend}
\end{figure}

In the inset to Fig.~\ref{FFT_TimeDepend}, we show the Fourier
transforms for the output and input signals, $F_1(\omega_0)$ and
$F_3(\omega_0)$, as a function of the total time $\omega_0 T/2
\pi$ when the system in the inset to Fig.~\ref{BandGap} with $N=30$ is driven 
at frequency $\omega_0=14.9$ and amplitude $A_0 = 10^{-6}$. We find only weak dependence of the Fourier transform on the
total time in the range $\omega_0 T/2\pi \gtrsim 10^{4.5}$.  In the
main panel of Fig.~\ref{FFT_TimeDepend}, we show that the gain
$G(\omega_0)$ versus pressure $p$ is nearly identical for $\omega_0
T/2\pi = 10^4$ and $10^5$. Thus, we selected $\omega_0 T/2\pi = 5.9
\times 10^4$ to calculate all of the discrete Fourier transforms. In
the inset to Fig.~\ref{FFT_DeltaDepend}, we show the dependence of the
Fourier transforms $F_1(\omega_0)$ and $F_3(\omega_0)$ on the sampling time $\omega_0 \Delta/2\pi$.  For
$\omega_0 \Delta/2\pi \lesssim 10^{-2}$, $F_1(\omega_0)$ and
$F_3(\omega_0)$ do not depend strongly on the sampling time. In the
main panel of Fig.~\ref{FFT_DeltaDepend}, we show that for most
pressures the gain $G(\omega_0)$ does not depend on $\Delta$. However,
at pressures for which there is large gain, we find that we need to
use $\omega_0 \Delta/2\pi \le 5.8 \times 10^{-3}$ to reach convergence.
Thus, we used this value of $\Delta$ to calculate all of the discrete
Fourier transforms.

\section*{Appendix B: Performance of Acoustic Switching Devices with Small Band Gaps}

\begin{figure}[h!]
\includegraphics[width=3in]{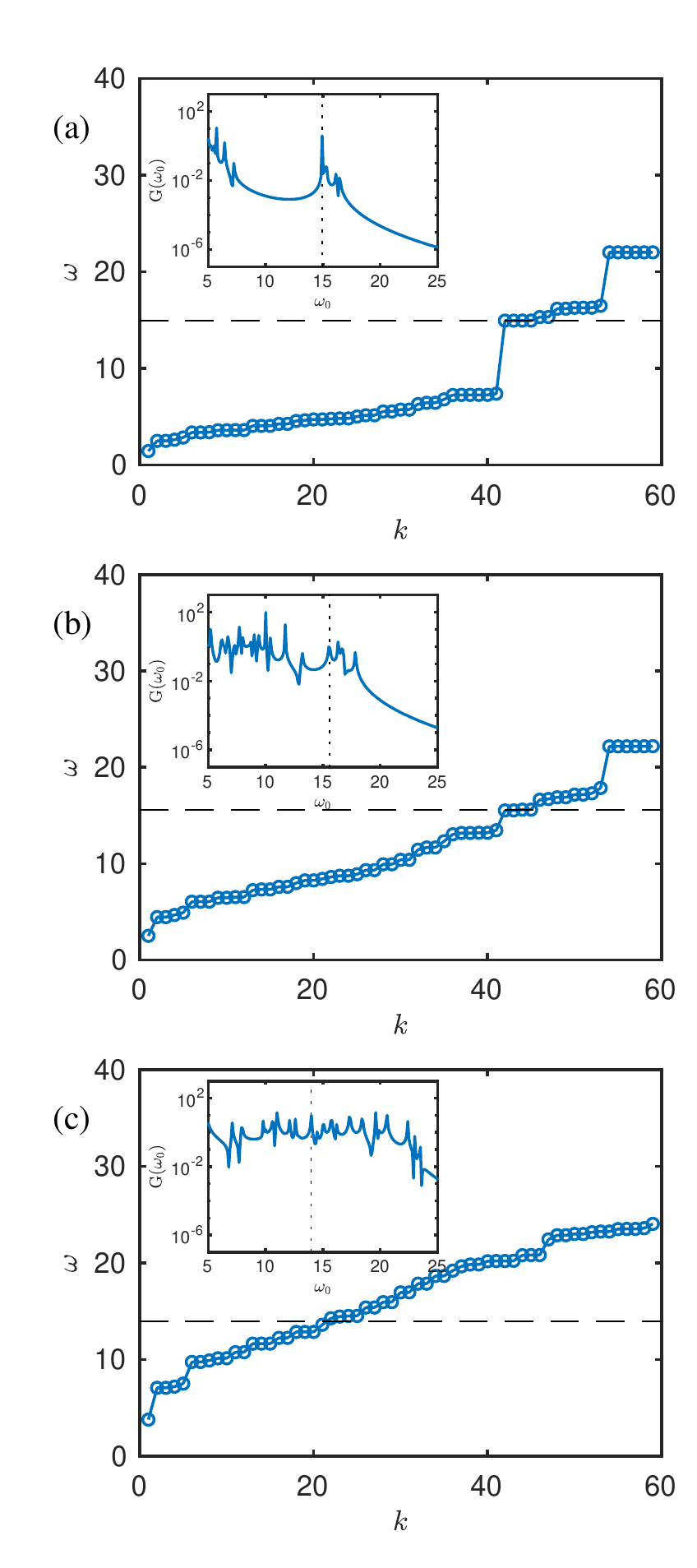}
\caption{Spectrum of eigenfrequencies for the mass-weighted dynamical 
matrix for the hexagonal lattice in the inset to Fig.~\ref{BandGap} with 
$N_L=21$ and $N_S=9$ for mass ratios (a) $m_L/m_S = 10$, (b) $3$, and (c) $1$.
The horizontal dashed lines indicate the frequencies at which we seek to 
drive the acoustic switching device. The insets of each panel show the frequency-dependent gain 
$G(\omega_0)$ (ratio 
of the Fourier transforms of the output and input signals) for the respective
mass ratios. For all systems, the pressure $p=10^{-1}$.}
\label{dGdp_mRatio_1}
\end{figure}

\begin{figure}[h!]
\includegraphics[width=3in]{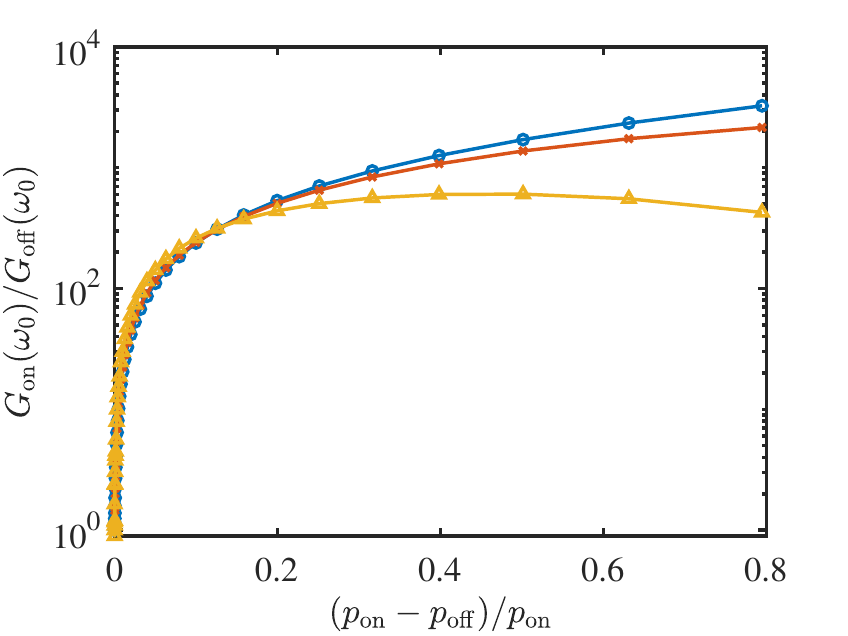}
\caption{The ratio of the gain $G_{\rm on}(\omega_0)/G_{\rm off}(\omega_0)$ in the on state to that in the off state
as a function of the normalized difference in pressure $(p_{\rm on}-p_{\rm off})/p_{\rm on}$ between the on and 
off states for mass ratios $m_L/m_S = 10$ (circles), $3$ (exes), and $1$ (triangles). The devices are driven at the frequencies $\omega_0$ indicated by 
the dashed lines in Fig.~\ref{dGdp_mRatio_1} (a)-(c).}
\label{dGdp_mRatio_2}
\end{figure}

In the main text, we described acoustic switching devices that possess
large frequency band gaps in their vibrational density of
states. However, we have not yet discussed how the performance of the
devices change with the size of the band gap.  In
Fig.~\ref{dGdp_mRatio_1}, we show the eigenfrequency spectrum of the
mass-weighted dynamical matrix for three mass ratios, $m_L/m_S=10$,
$3$, and $1$.  As shown previously in Fig.~\ref{BandGap}, the maximum
difference between adjacent eigenfrequencies $w$ decreases as $m_L/m_S
\rightarrow 1$. When there is a large band gap, we expect that we can
generate a well-defined on state by driving the system at an
eigenfrequency that populates the vibrational density of states.  In
addition, we expect that we can create a well-defined off state,
e.g. by decreasing the pressure of the system (which increases all of
the eigenfrequencies) so that the driving frequency now occurs within
the band gap. (See the difference between the exes and circles in
Fig.~\ref{AllPart_w}.)  The frequency-dependent gain (ratio of the
Fourier transforms of the output to the input signal) for a system
with a large band gap is shown in the inset to Fig.~\ref{dGdp_mRatio_1}
(a). Indeed, the gain at $\omega=14.9$ is $\approx 10$, while the gain
at nearby \textit{lower} frequencies is several orders of magnitude
lower. Thus, it is clear that an acoustic switch can be created by
choosing the on state as the system with reference pressure
($p=10^{-1}$) in Fig.~\ref{dGdp_mRatio_1} (a) driven at frequency
$\omega_0=14.9$ and choosing the pressure for the off state so that
$G(\omega_0) \lesssim 10^{-2}$. Similar behavior is shown in
Fig.~\ref{dGdp_mRatio_1} (b) for a system with a smaller band gap at
mass ratio $m_L/m_S=3$.  For example, the on state can be generated by
driving the system at $\omega_0=15.5$, where the gain possesses a
peak. The gain at nearby lower frequencies is smaller, but the gain
has another peak at $\omega_0=13.2$. Thus, the operating range of the
pressure difference of the acoustic switch decreases as the band gap
decreases.

We now focus on the continuous eigenfrequency regime near $\omega_0
=14.0$ for systems with $m_L/m_S=1$ in Fig.~\ref{dGdp_mRatio_1} (c). The
frequency-dependent gain possesses a peak at $\omega_0=14.0$, but the
next peak in $G(\omega_0)$ at lower frequency does not occur until
$\omega_0 =12.6$, even though the eigenfrequency spectrum includes $3$
eigenfrequencies between $12.6$ and $14.0$.  For these
eigenfrequencies, the overlap between the eigenmodes and either the
input or output signal is small, and thus the output signal is weak
when the system is driven at these eigenfrequencies. As a result, the
gain ratio can be large even for systems with a continuous
eigenfrequency spectrum. In Fig.~\ref{dGdp_mRatio_2}, we show that the
device with $m_L/m_S=1$ can achieve a gain ratio $G_{\rm
  on}(\omega_0)/G_{\rm off}(\omega_0) > 10^2$.

Thus, we have shown that the vibrational response of the device at a
given eigenfrequency depends on the overlap between the eigenmodes near 
the driving frequency and the input and output signals.  A robust  
acoustic switch can always be produced using a system with a finite
frequency band gap.  However, an acoustic switch can also be created
using a system with a continuous eigenfrequency spectrum if the
driving frequency is chosen such that the eigenmodes of the
corresponding nearby eigenfrequencies do not couple to the input and
output signals.  Such acoustic switching devices are more difficult to
design since one needs to control the spatial structure of the
eigenmodes, as well as the eigenfrequency spectrum.

\section*{Acknowledgments}

The authors acknowledge financial support from NSF Grant
Nos. CMMI-1462439 (C.O. and Q.W.), CMMI-1463455 (M.S.), and
CBET-1605178 (C.O. and Q.W.).  We also acknowledge Tsinghua University
that supported Chunyang Cui's visit to Yale University and the Kavli
Institute for Theoretical Physics (under NSF Grant No. PHY-1748958),
where this work was completed. In addition, this work was supported by
the High Performance Computing facilities operated by, and the staff
of, the Yale Center for Research Computing.

\bibliography{qikai.bib}

\end{document}